\pdfoutput=1
\documentclass[12pt,a4paper]{article}

\usepackage{float}
\usepackage{placeins}

\usepackage{ifthen} 
\newboolean{pdflatex}
\setboolean{pdflatex}{true} 

\newboolean{articletitles}
\setboolean{articletitles}{true} 

\newboolean{uprightparticles}
\setboolean{uprightparticles}{false} 

\def\hm{\ensuremath{h^-}\xspace}
\def\hp{\ensuremath{h^+}\xspace}

\def\paperauthors{LHCb collaboration} 
\def\paperasciititle{Constraints on the CKM angle gamma from B->Dh decays using  D->hhpi0 final states} 
\def\papertitle{Constraints on the CKM angle $\gamma$ from $B^\pm\rightarrow Dh^\pm$ decays\\ using  $D\rightarrow h^\pm h^{\prime\mp}\piz$ final states} 
\def\paperkeywords{{High Energy Physics}, {LHCb}} 
\def\papercopyright{\the\year\ CERN for the benefit of the LHCb collaboration} 
\def\paperlicence{CC BY 4.0 licence}
\def\paperlicenceurl{https://creativecommons.org/licenses/by/4.0/}

\usepackage[top=1in, bottom=1.25in, left=1in, right=1in]{geometry}

\usepackage{microtype}
\usepackage{lineno}  
\usepackage{xspace} \usepackage{caption}

\usepackage{subcaption} 
\usepackage{lscape} 

\usepackage{graphicx}  
\usepackage{color}
\usepackage{colortbl}
\graphicspath{{./figs/}} 

\usepackage{amsmath} 
\usepackage{amssymb}
\usepackage{amsfonts}
\usepackage{upgreek} 

\newcommand*\patchAmsMathEnvironmentForLineno[1]{%
\expandafter\let\csname old#1\expandafter\endcsname\csname #1\endcsname
\expandafter\let\csname oldend#1\expandafter\endcsname\csname
end#1\endcsname
 \renewenvironment{#1}%
   {\linenomath\csname old#1\endcsname}%
   {\csname oldend#1\endcsname\endlinenomath}%
}
\newcommand*\patchBothAmsMathEnvironmentsForLineno[1]{%
  \patchAmsMathEnvironmentForLineno{#1}%
  \patchAmsMathEnvironmentForLineno{#1*}%
}
\AtBeginDocument{%
\patchBothAmsMathEnvironmentsForLineno{equation}%
\patchBothAmsMathEnvironmentsForLineno{align}%
\patchBothAmsMathEnvironmentsForLineno{flalign}%
\patchBothAmsMathEnvironmentsForLineno{alignat}%
\patchBothAmsMathEnvironmentsForLineno{gather}%
\patchBothAmsMathEnvironmentsForLineno{multline}%
\patchBothAmsMathEnvironmentsForLineno{eqnarray}%
}

\usepackage{hyperxmp}

\usepackage[pdftex,
            pdfauthor={\paperauthors},
            pdftitle={\paperasciititle},
            pdfkeywords={\paperkeywords},
            pdfcopyright={Copyright (C) \papercopyright},
            pdflicenseurl={\paperlicenceurl}]{hyperref}

\usepackage[colorinlistoftodos,textsize=scriptsize]{todonotes}

\usepackage[bottom,flushmargin,hang,multiple]{footmisc}

\usepackage[all]{hypcap} 

\usepackage{xspace} 
\usepackage{upgreek}

\def\lhcb   {\mbox{LHCb}\xspace}

\def\MagUp {\mbox{\em Mag\kern -0.05em Up}\xspace}

\ifthenelse{\boolean{uprightparticles}}%
{
 
 \def\Pgamma      {\ensuremath{\upgamma}\xspace}

 \def\Ppi         {\ensuremath{\uppi}\xspace}

 \def\PDelta      {\ensuremath{\Delta}\xspace}                 
 \def\PXi         {\ensuremath{\Xi}\xspace}                 
 \def\PLambda     {\ensuremath{\Lambda}\xspace}                 
 \def\PSigma      {\ensuremath{\Sigma}\xspace}                 
 \def\POmega      {\ensuremath{\Omega}\xspace}                 
 \def\PUpsilon    {\ensuremath{\Upsilon}\xspace}

 \def\PB      {\ensuremath{\mathrm{B}}\xspace}                 
                  
 \def\PD      {\ensuremath{\mathrm{D}}\xspace}

 \def\PK      {\ensuremath{\mathrm{K}}\xspace}

 \def\Pb      {\ensuremath{\mathrm{b}}\xspace}                 
 \def\Pc      {\ensuremath{\mathrm{c}}\xspace}

 \def\Ph      {\ensuremath{\mathrm{h}}\xspace}                 
 \def\Pi      {\ensuremath{\mathrm{i}}\xspace}

 \def\Ps      {\ensuremath{\mathrm{s}}\xspace}                 
                  
 \def\Pu      {\ensuremath{\mathrm{u}}\xspace}

 \def\thebaroffset{0.0em}
}
{
 
 \def\Pgamma      {\ensuremath{\gamma}\xspace}

 \def\Ppi         {\ensuremath{\pi}\xspace}

 \mathchardef\PDelta="7101
 \mathchardef\PXi="7104
 \mathchardef\PLambda="7103
 \mathchardef\PSigma="7106
 \mathchardef\POmega="710A
 \mathchardef\PUpsilon="7107
                  
 \def\PB      {\ensuremath{B}\xspace}                 
                  
 \def\PD      {\ensuremath{D}\xspace}

 \def\PK      {\ensuremath{K}\xspace}

 \def\Pb      {\ensuremath{b}\xspace}                 
 \def\Pc      {\ensuremath{c}\xspace}

 \def\Ph      {\ensuremath{h}\xspace}                 
 \def\Pi      {\ensuremath{i}\xspace}

 \def\Ps      {\ensuremath{s}\xspace}                 
                  
 \def\Pu      {\ensuremath{u}\xspace}

 \def\thebaroffset{0.18em}
}
\newcommand{\offsetoverline}[2][\thebaroffset]{\kern #1\overline{\kern -#1 #2}}%

\makeatletter
\ifcase \@ptsize \relax
  \newcommand{\miniscule}{\@setfontsize\miniscule{4}{5}}
\or
  \newcommand{\miniscule}{\@setfontsize\miniscule{5}{6}}
\or
  \newcommand{\miniscule}{\@setfontsize\miniscule{5}{6}}
\fi
\makeatother

\DeclareRobustCommand{\optbar}[1]{\shortstack{{\miniscule (\rule[.5ex]{1.25em}{.18mm})}
  \\ [-.7ex] $#1$}}

\def\g      {{\ensuremath{\Pgamma}}\xspace}

\def\uquark    {{\ensuremath{\Pu}}\xspace}
\def\uquarkbar {{\ensuremath{\overline \uquark}}\xspace}

\def\squark    {{\ensuremath{\Ps}}\xspace}

\def\cquark    {{\ensuremath{\Pc}}\xspace}
\def\cquarkbar {{\ensuremath{\overline \cquark}}\xspace}

\def\bquark    {{\ensuremath{\Pb}}\xspace}

\def\hadron {{\ensuremath{\Ph}}\xspace}
\def\pion   {{\ensuremath{\Ppi}}\xspace}
\def\piz    {{\ensuremath{\pion^0}}\xspace}
\def\pip    {{\ensuremath{\pion^+}}\xspace}
\def\pim    {{\ensuremath{\pion^-}}\xspace}
\def\pipm   {{\ensuremath{\pion^\pm}}\xspace}
\def\pimp   {{\ensuremath{\pion^\mp}}\xspace}

\def\kaon    {{\ensuremath{\PK}}\xspace}

\def\KorKbar {\kern \thebaroffset\optbar{\kern -\thebaroffset \PK}{}\xspace}

\def\Kp      {{\ensuremath{\kaon^+}}\xspace}
\def\Km      {{\ensuremath{\kaon^-}}\xspace}
\def\Kpm     {{\ensuremath{\kaon^\pm}}\xspace}
\def\Kmp     {{\ensuremath{\kaon^\mp}}\xspace}

\def\Kstar   {{\ensuremath{\kaon^*}}\xspace}

\def\Dbar    {{\ensuremath{\offsetoverline{\PD}}}\xspace}
\def\D       {{\ensuremath{\PD}}\xspace}

\def\DorDbar {\kern \thebaroffset\optbar{\kern -\thebaroffset \PD}\xspace}
\def\Dz      {{\ensuremath{\D^0}}\xspace}
\def\Dzb     {{\ensuremath{\Dbar{}^0}}\xspace}
\def\Dp      {{\ensuremath{\D^+}}\xspace}
\def\Dm      {{\ensuremath{\D^-}}\xspace}

\def\DpDm    {\ensuremath{\Dp {\kern -0.16em \Dm}}\xspace}
\def\Dstar   {{\ensuremath{\D^*}}\xspace}

\def\Dstarz  {{\ensuremath{\D^{*0}}}\xspace}

\def\Dstarpm {{\ensuremath{\D^{*\pm}}}\xspace}

\def\B       {{\ensuremath{\PB}}\xspace}

\def\BorBbar {\kern \thebaroffset\optbar{\kern -\thebaroffset \PB}\xspace}

\def\Bd      {{\ensuremath{\B^0}}\xspace}

\def\BdorBdbar {\kern \thebaroffset\optbar{\kern -\thebaroffset \Bd}\xspace}
\def\Bu      {{\ensuremath{\B^+}}\xspace}
\def\Bub     {{\ensuremath{\B^-}}\xspace}
\def\Bp      {{\ensuremath{\Bu}}\xspace}
\def\Bm      {{\ensuremath{\Bub}}\xspace}
\def\Bpm     {{\ensuremath{\B^\pm}}\xspace}
\def\Bmp     {{\ensuremath{\B^\mp}}\xspace}
\def\Bs      {{\ensuremath{\B^0_\squark}}\xspace}

\def\BsorBsbar {\kern \thebaroffset\optbar{\kern -\thebaroffset \Bs}\xspace}

\def\Y#1S{\ensuremath{\PUpsilon{(#1S)}}\xspace}

\def\LorLbar     {\kern \thebaroffset\optbar{\kern -\thebaroffset \PLambda}\xspace}

\def\to                 {\ensuremath{\rightarrow}\xspace}

\def\CP                {{\ensuremath{C\!P}}\xspace}

\def\AT#1     {\ensuremath{A_{\mathrm{T}}^{#1}}\xspace}           

\def\C#1      {\ensuremath{\mathcal{C}_{#1}}\xspace}                       
\def\Cp#1     {\ensuremath{\mathcal{C}_{#1}^{'}}\xspace}                    
\def\Ceff#1   {\ensuremath{\mathcal{C}_{#1}^{\mathrm{(eff)}}}\xspace}        
\def\Cpeff#1  {\ensuremath{\mathcal{C}_{#1}^{'\mathrm{(eff)}}}\xspace}       
\def\Ope#1    {\ensuremath{\mathcal{O}_{#1}}\xspace}                       
\def\Opep#1   {\ensuremath{\mathcal{O}_{#1}^{'}}\xspace}                    

\newcommand{\nospaceunit}[1]{\ensuremath{\text{#1}}}       
\newcommand{\aunit}[1]{\ensuremath{\text{\,#1}}}       

\newcommand{\tev}{\aunit{Te\kern -0.1em V}\xspace}
\newcommand{\gev}{\aunit{Ge\kern -0.1em V}\xspace}
\newcommand{\mev}{\aunit{Me\kern -0.1em V}\xspace}
\newcommand{\kev}{\aunit{ke\kern -0.1em V}\xspace}
\newcommand{\ev}{\aunit{e\kern -0.1em V}\xspace}
 
\newcommand{\mevc}{\ensuremath{\aunit{Me\kern -0.1em V\!/}c}\xspace}
\newcommand{\gevc}{\ensuremath{\aunit{Ge\kern -0.1em V\!/}c}\xspace}
\newcommand{\mevcc}{\ensuremath{\aunit{Me\kern -0.1em V\!/}c^2}\xspace}
\newcommand{\gevcc}{\ensuremath{\aunit{Ge\kern -0.1em V\!/}c^2}\xspace}

\def\mum  {\ensuremath{\,\upmu\nospaceunit{m}}\xspace}

\def\fb   {\ensuremath{\aunit{fb}}\xspace}
\def\invfb   {\ensuremath{\fb^{-1}}\xspace}

\newcommand{\chisq}{\ensuremath{\chi^2}\xspace}

\newcommand{\chisqip}{\ensuremath{\chi^2_{\text{IP}}}\xspace}

\def\gsim{{~\raise.15em\hbox{$>$}\kern-.85em
          \lower.35em\hbox{$\sim$}~}\xspace}
\def\lsim{{~\raise.15em\hbox{$<$}\kern-.85em
          \lower.35em\hbox{$\sim$}~}\xspace}

\def\pt         {\ensuremath{p_{\mathrm{T}}}\xspace}

\def\ptot       {\ensuremath{p}\xspace}

\def\evtgen     {\mbox{\textsc{EvtGen}}\xspace}

\def\geant      {\mbox{\textsc{Geant4}}\xspace}

\def\photos     {\mbox{\textsc{Photos}}\xspace}

\def\pythia     {\mbox{\textsc{Pythia}}\xspace}

\def\tell1  {TELL1\xspace}
\def\ukl1   {UKL1\xspace}

\newcommand{\eg}{\mbox{\itshape e.g.}\xspace}

\usepackage{cite} 
\usepackage{mciteplus}

\begin{document}

\renewcommand{\thefootnote}{\fnsymbol{footnote}}
\setcounter{footnote}{1}

\begin{titlepage}
\pagenumbering{roman}

\vspace*{-1.5cm}
\centerline{\large EUROPEAN ORGANIZATION FOR NUCLEAR RESEARCH (CERN)}
\vspace*{1.5cm}
\noindent
\begin{tabular*}{\linewidth}{lc@{\extracolsep{\fill}}r@{\extracolsep{0pt}}}
\ifthenelse{\boolean{pdflatex}}
{\vspace*{-1.5cm}\mbox{\!\!\!\includegraphics[width=.14\textwidth]{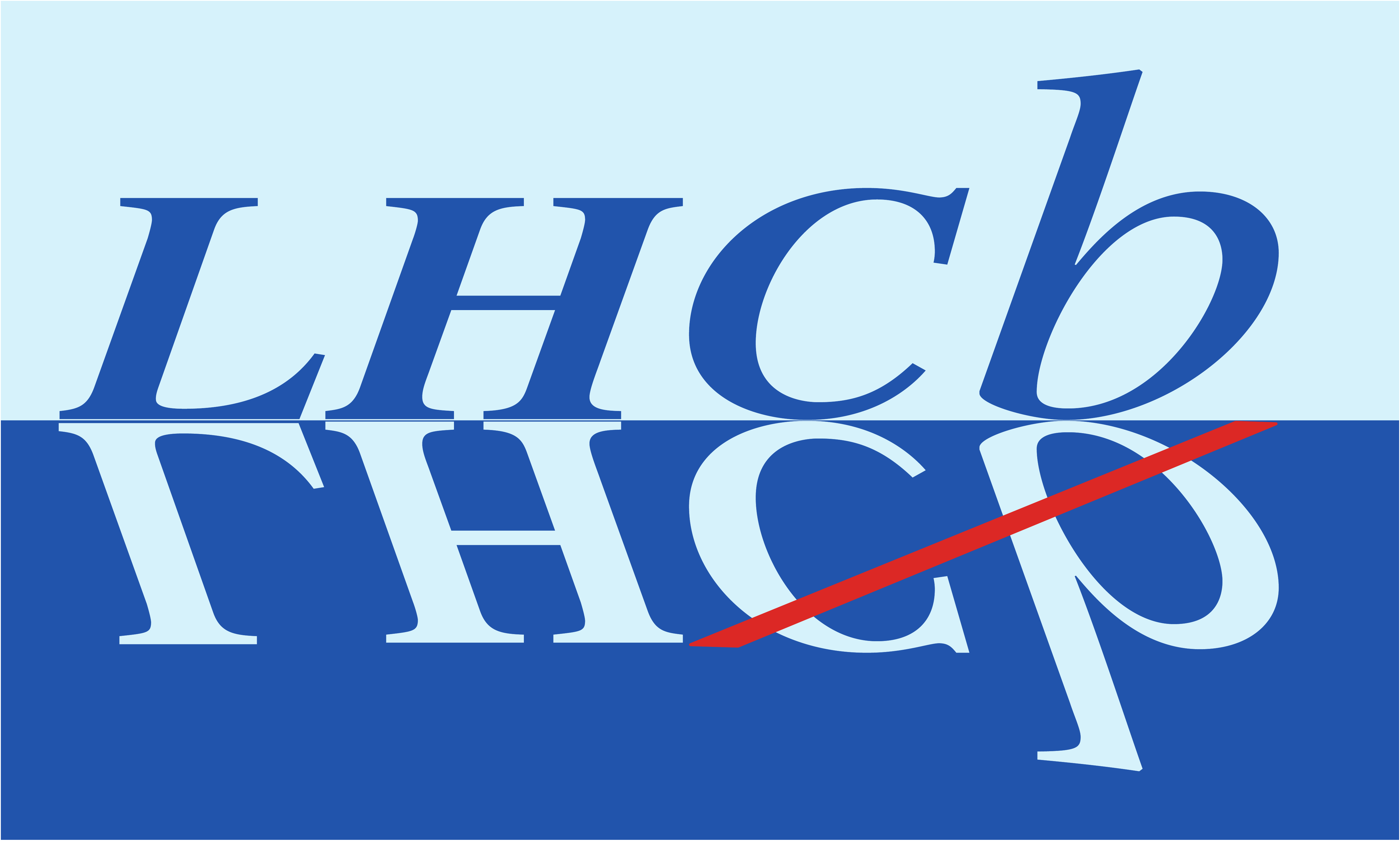}} & &}%
{\vspace*{-1.2cm}\mbox{\!\!\!\includegraphics[width=.12\textwidth]{lhcb-logo.eps}} & &}%
\\
 & & CERN-EP-2021-255 \\  
 & & LHCb-PAPER-2021-036 \\  
 & & 21 July 2022 \\ 
 & & \\
\end{tabular*}

\vspace*{2.0cm}

{\normalfont\bfseries\boldmath\huge
\begin{center}
  \papertitle 
\end{center}
}

\vspace*{2.0cm}

\begin{center}
\paperauthors\footnote{Authors are listed at the end of this paper.}
\end{center}

\vspace*{0.3cm}

\begin{center}
 \textit{ This paper is dedicated to our friend and colleague Bernhard Spaan.}  
\end{center}

\vspace*{1.0cm}

\vspace{\fill}

\begin{abstract}
  \noindent
  A data sample collected with the LHCb detector corresponding to an integrated luminosity of $9~\invfb$ is used to measure eleven \CP violation observables in \mbox{$\Bpm \to \D \hadron^\pm$} decays, where $h$ is either a kaon or a pion. The neutral $\D$ meson decay is reconstructed in the three-body final states: \mbox{$\Kpm\pimp\piz$}; \mbox{$\pip\pim\piz$}; \mbox{$\Kp\Km\piz$} and the suppressed \mbox{$\pipm\Kmp\piz$} combination. The mode where a large \CP asymmetry is expected, \mbox{$\Bpm\to [\pipm\Kmp\piz]_D\Kpm$}, is observed with a significance greater than seven standard deviations. The ratio of the partial width of this mode relative to that of the favoured mode, \mbox{$\Bpm\to [\Kpm\pimp\piz]_D\Kpm$}, is
  \mbox{$R_{{\rm ADS}(K)}=(1.27\pm0.16\pm0.02)\times 10^{-2}$}. Evidence for a large \CP asymmetry is also seen: \mbox{$A_{{\rm ADS}(K)}=-0.38\pm0.12\pm0.02$}.   Constraints on the CKM angle $\gamma$ are calculated from the eleven reported observables.
  \end{abstract}

\vspace*{1.0cm}

\begin{center}
  Published in JHEP 07 (2022) 099 
\end{center}

\vspace{\fill}

{\footnotesize 
\centerline{\copyright~\papercopyright. \href{\paperlicenceurl}{\paperlicence}.}}
\vspace*{2mm}

\end{titlepage}
\newpage
\setcounter{page}{2}
\mbox{~}

\renewcommand{\thefootnote}{\arabic{footnote}}
\setcounter{footnote}{0}

\cleardoublepage

\pagestyle{plain} 
\setcounter{page}{1}
\pagenumbering{arabic}


\newcommand{\asymmTerm}[3]{ \Gamma(#1)#2\Gamma(#3) }
\newcommand{\asymmEq}[5]{
\begin{equation}
    \label{eq:#4}
    #1=\frac{\asymmTerm{#2}{-}{#3}}{\asymmTerm{#2}{+}{#3}}#5
\end{equation}
}
\newcommand{\Bdecaych}[3]{\mbox{\ensuremath{\Bm\rightarrow [#1#2\piz]_\D#3}}}
\newcommand{\hFAV}{\Bdecaych{\Km}{\pip}{\hm}\xspace}
\newcommand{\piFAV}{\Bdecaych{\Km}{\pip}{\pim}\xspace}
\newcommand{\kSUP}{\Bdecaych{\pim}{\Kp}{\Km}\xspace}
\newcommand{\hGLWK}{\Bdecaych{\Km}{\Kp}{\hm}\xspace}
\newcommand{\hGLWPi}{\Bdecaych{\pim}{\pip}{\hm}\xspace}
\newcommand{\kGLWPi}{\Bdecaych{\pim}{\pip}{\Km}\xspace}
\newcommand{\btodh}{\mbox{\ensuremath{\Bm\to\D\hm}}\xspace}
\newcommand{\btodk}{\mbox{\ensuremath{\Bm\to\D\Km}}\xspace}
\newcommand{\btodpi}{\mbox{\ensuremath{\Bm\to\D\pim}}\xspace}
\newcommand{\Aprod}{A_{\rm Prod}}
\newcommand{\kFAV}{\ensuremath{\Bm\to[\Km\pip\piz]_D\Km}\xspace}
\newcommand{\kFAVpm}{\ensuremath{\Bmp\to[\Kmp\pipm\piz]_D\Kmp}\xspace}
\newcommand{\kSUPpm}{\ensuremath{\Bmp\to[\pimp\Kpm\piz]_D\Kmp}\xspace}
\newcommand{\piFAVpm}{\ensuremath{\Bmp\to[\Kmp\pipm\piz]_D\pimp}\xspace}
\newcommand{\piSUPpm}{\ensuremath{\Bmp\to[\pimp\Kpm\piz]_D\pimp}\xspace}
\newcommand{\piSUP}{\ensuremath{\Bm\to[\pim\Kp\piz]_D\pim}\xspace}
\newcommand{\hSUP}{\ensuremath{\Bm\to[\pim\Kp\piz]_D\hm}\xspace}
\newcommand{\ratioFrac}[2]{\frac{\Gamma(#1)}{\Gamma(#2)}}
\newcommand{\BminusDecay}[3]{\Bm\to[#1#2\piz]_D#3}
\newcommand{\BplusDecay}[3]{\Bp\to[#1#2\piz]_D#3}
\newcommand{\kFAVminus}{ \BminusDecay{\Km}{\pip}{\Km} }
\newcommand{\kFAVplus}{ \BplusDecay{\Kp}{\pim}{\Kp} }
\newcommand{\piFAVminus}{ \BminusDecay{\Km}{\pim}{\pim} }
\newcommand{\piFAVplus}{ \BplusDecay{\Kp}{\pim}{\pip} }
\newcommand{\kSUPminus}{ \BminusDecay{\pim}{\Kp}{\Km} }
\newcommand{\kSUPplus}{ \BplusDecay{\pip}{\Km}{\Kp} }
\newcommand{\piSUPminus}{ \BminusDecay{\pim}{\Kp}{\pim} }
\newcommand{\piSUPplus}{ \BplusDecay{\pip}{\Km}{\pip} }
\newcommand{\Akfav}{ A_\kaon^{\kaon\pi\piz} }
\newcommand{\Apifav}{ A_\pi^{\kaon\pi\piz} }
\newcommand{\Akglwk}{ A_\kaon^{\kaon\kaon\piz} }
\newcommand{\Apiglwk}{ A_\pi^{\kaon\kaon\piz} }
\newcommand{\Akglwpi}{ A_\kaon^{\pi\pi\piz} }
\newcommand{\Apiglwpi}{ A_\pi^{\pi\pi\piz} }
\newcommand{\rkplus}{ R_\kaon^+ }
\newcommand{\rkminus}{ R_\kaon^- }
\newcommand{\rpiplus}{ R_\pi^+ }
\newcommand{\rpiminus}{ R_\pi^- }
\newcommand{\rb}{r_\B}
\newcommand{\rbk}{\ensuremath{r_\B^\kaon}\xspace}
\newcommand{\rbpi}{\ensuremath{r_\B^\pi}\xspace}
\newcommand{\db}{\delta_\B}
\newcommand{\dbk}{\ensuremath{\delta_\B^\kaon}\xspace}
\newcommand{\dbpi}{\delta_\B^\pi}
\newcommand{\kpp}{\kaon\pi\piz}
\newcommand{\pkp}{\pi\kaon\piz}
\newcommand{\kkp}{\kaon\kaon\piz}
\newcommand{\ppp}{\pi\pi\piz}
\newcommand{\kapfav}{\kappa^{\kaon\pi\piz}_D}
\newcommand{\delfav}{\delta_D^{\kaon\pi\piz}}
\newcommand{\effkk}{F_+^{\kaon\kaon\piz}}
\newcommand{\effpp}{F_+^{\pi\pi\piz}}
\newcommand{\rdfav}{r_D^{\kpp}}
\newcommand{\hprime}{h^\prime}
\newcommand{\effhh}{F_+^{\hprime\hprime\piz}}
\newcommand{\Rglw}{R^{hh\piz}}
\newcommand{\Rglwkk}{R^{KK\piz}}
\newcommand{\Rglwpipi}{R^{\pi\pi\piz}}
\newcommand{\Rkpihh}{R^{hh\piz}_{\kaon/\pi}}
\newcommand{\Rkpikk}{R^{KK\piz}_{\kaon/\pi}}
\newcommand{\Rkpipipi}{R^{\pi\pi\piz}_{\kaon/\pi}}
\newcommand{\Aglw}{A^{hh\piz}_{h}}
\newcommand{\Aglwk}{A^{hh\piz}_{K}}
\newcommand{\Aglwpi}{A^{hh\piz}_{\pi}}
\newcommand{\Rkpi}{R^{\kaon\pi\piz}_{\kaon/\pi}}
\newcommand{\hhp}{hh\piz}
\newcommand{\kppQ}{\Kp\pim\piz}
\newcommand{\BmtoGLWh}{\Bm\to [\hhp]_D\hm}
\newcommand{\BptoGLWh}{\Bp\to [\hhp]_D\hp}
\newcommand{\BmtoGLWk}{\Bm\to [\hhp]_D\Km}
\newcommand{\BptoGLWk}{\Bp\to [\hhp]_D\Kp}
\newcommand{\BmtoGLWp}{\Bm\to [\hhp]_D\pim}
\newcommand{\BptoGLWp}{\Bp\to [\hhp]_D\pip}
\newcommand{\BmtoFAVh}{\Bm\to [\kppQ]_D\hm}
\newcommand{\BptoFAVh}{\Bp\to [\kppQ]_D\hp}
\newcommand{\BmtoFAVk}{\Bm\to [\kppQ]_D\Km}
\newcommand{\BptoFAVk}{\Bp\to [\kppQ]_D\Kp}
\newcommand{\BmtoFAVp}{\Bm\to [\kppQ]_D\pim}
\newcommand{\BptoFAVp}{\Bp\to [\kppQ]_D\pip}

\section{Introduction}
\label{sec:intro}
Precision measurements of the Cabibbo-Kobayashi-Maskawa (CKM) unitarity triangle parameters are essential in the search for new physics in the quark flavour sector.
The unitarity triangle can be overconstrained by measuring its three angles and two lengths.
The angle $\gamma\equiv\arg(-V^{\phantom{*}}_{ud}V^*_{ub}/V^{\phantom{*}}_{cd}V^*_{cb})$, where $V_{ik}$ are the CKM matrix elements, is an ideal Standard Model benchmark since it is independent of top-quark couplings.
It can be determined from measurements of \CP violation in decays that are dominated by tree-level contributions with negligible theoretical uncertainties \cite{Brod:2013sga}.
The world-average value, $\gamma=(66.2^{+3.4}_{-3.6})^\circ$~\cite{HFLAV18}, is in good agreement with the value inferred from a global CKM fit where direct $\gamma$ determinations are excluded: $\gamma=(65.5^{+1.1}_{-2.7})^\circ$~\cite{CKMfitter2005}. To test this agreement at the sub-degree level, it is important to develop new modes to complement established techniques.

\newcommand{\DDec}{\Bm\to\Dz\Km}
\newcommand{\DbarDec}{\Bm\to\Dzb\Km}
\newcommand{\bToDK}{\Bm\to\D\Km}
\newcommand{\bToDPi}{\Bm\to\D\pim}

The angle $\gamma$ is the relative weak phase between $\bquark\to\cquark\uquarkbar\squark$ and $\bquark\to\uquark\cquarkbar\squark$ quark transition amplitudes.
These transitions mediate the decays $\Bm\to\Dz\Km$ and $\Bm\to\Dzb\Km$, respectively.\footnote{The inclusion of charge-conjugate modes is implied everywhere except when discussing asymmetries.}
By studying final states accessible to both $\Dz$ and $\Dzb$ mesons, phase information can be determined from the interference of the two amplitudes.
As well as \g, the ratio of the magnitudes of the $\DbarDec$ to $\DDec$ amplitudes, $r_B\approx0.1$, and the relative strong phase, $\delta_\B$, determine the size of the interference.
Such interference also occurs in $\bToDPi$ decays, albeit with lower \g sensitivity due to additional Cabibbo suppression that leaves the amplitude ratio around $20$ times smaller than for \bToDK decays. Here \D represents an admixture of the \Dz and \Dzb states.

The measurement of $\gamma$ through $\bToDK$ decays was first suggested by Gronau, London \& Wyler for \D decays reconstructed in \CP eigenstates; they are often referred to as GLW modes \cite{GRONAU1991483, GRONAU1991172}.
The method was generalised by Atwood, Dunietz \& Soni (ADS) to include non-charge-conjugate states \cite{Atwood:1996ci,PhysRevD.63.036005}.
In ADS modes, the favoured (suppressed) \btodh decay is followed by a suppressed (favoured) \D meson decay, which has the effect of roughly balancing the two competing amplitudes, maximising their interference and thus their sensitivity to \g. 
There are also favoured decays, where the $\bquark\to\cquark\uquarkbar\squark$ transition is followed by a favoured \D meson decay. These have little interference and weak sensitivity to \g, but provide appropriate normalisation for the suppressed decays.

A search for $\bquark\to\uquark\cquarkbar\squark$ amplitudes contributing to \btodk decays was first performed with the \mbox{$D\to K\pi\pi^0$} mode by the BaBar collaboration~\cite{BaBar:2011rud}. Later, evidence of the suppressed \kSUP decay was reported by the BELLE collaboration~\cite{Belle:2013dtr} and \lhcb~\cite{LHCb-PAPER-2015-014}. This work supersedes Ref.~\cite{LHCb-PAPER-2015-014} with a fourfold increase in data. 


\section{External inputs and formalism}
\label{sec:formalism}

Three-body GLW modes, like those considered in this paper: \mbox{$\D\to\pim\pip\piz$} and \mbox{$\D\to\Km\Kp\piz$}, are an admixture of \CP-even and \CP-odd eigenstates.
\CP-even and \CP-odd states exhibit opposite \CP asymmetry so integrating over the three-body phase space dilutes sensitivity to \g.
This dilution is parameterised by $\CP$-even fractions: \mbox{$F^{\pi\pi\piz}_{+}$} and \mbox{$F^{KK\piz}_{+}$}. A $\CP$-even fraction of zero implies a pure $\CP$-odd state, 0.5 means an equal amount of $\CP$-even and $\CP$-odd states contributing to the multibody decay. The $\CP$-even fractions relevant to this analysis have been measured with quantum-correlated $\D\Dbar$ pairs produced at the $\psi(3770)$ resonance: $F^{\pi\pi\piz}_{+}=0.973\pm0.017$ and $F^{KK\piz}_{+}=0.732\pm0.055$ are measured in Refs.~\cite{Malde:2015mha,Nayak:2014tea}.

In the case of the non-charge-conjugate ADS mode, both the Cabibbo-suppressed $\Dzb\to\Km\pip\piz$ and favoured $\Dz\to\Km\pip\piz$ amplitudes contribute. The ratio of their magnitudes, $r_D=0.0441\pm0.0011$~\cite{Ablikim:2021cqw} is similar to $r_B$, hence the possibility of large \CP asymmetries. In the ADS case, the dilution is parameterised by a coherence factor, $\kappa_D=0.79\pm0.04$. This number and the average strong phase difference between $\Dzb\to\Km\pip\piz$ and $\Dz\to\Km\pip\piz$ amplitudes, $\delta_D=(196\pm11)^\circ$, are reported in Ref.~\cite{Ablikim:2021cqw}. 
The relatively high value of $\kappa_D$ means that integrating over the whole three-body \D decay phase space retains sensitivity to \g.

Incorporating the effect of \D meson mixing up to first order in $x$ and $y$~\cite{Rama:2013voa}, the partial rate expression  for the ADS modes is
\newcommand{\rdf}{r_D^f}
\newcommand{\rbx}{r_B^X}
\newcommand{\ddf}{\delta_D^f}
\newcommand{\dbx}{\delta_B^X}
\begin{eqnarray}
\!\!\Gamma(\B^\mp\to[\pi^\mp K^\pm\piz]_D h^\mp) &\propto& r_D^2+r_B^2+2r_D r_B\kappa_D\cos(\delta_B+\delta_D\mp\gamma) \nonumber\\
&& -\,\alpha y(1+r_B^2)r_D\kappa_D\cos\delta_D -\alpha y(1+r_D^2)r_B\cos(\delta_B\mp\gamma)\ \ \  \label{rateSup}  \\
&&+\, \alpha x(1-r_B^2)r_D\kappa_D\sin\delta_D-\alpha x(1-r_D^2)r_B\sin(\delta_B\mp\gamma), \nonumber  
\end{eqnarray}
where $x=(0.409^{+0.048}_{-0.049})\%$ and $y=(0.615^{+0.056}_{-0.055})\%$ are the charm mixing parameters~\cite{HFLAV18} and $\alpha$ is an analysis-specific coefficient that quantifies the decay-time acceptance of the candidate $D$ mesons.
This coefficient is determined from simulation, to be $\alpha=1.0$ with negligible uncertainty.
The same rate expression is used for GLW mode $f$, where $r_D\equiv1$, $\delta_D\equiv0$ and $\kappa_D=2F_+^f-1$.
For completeness, the equivalent rate for the favoured mode is
\begin{eqnarray}
\!\!\Gamma(\B^\mp\to[K^\mp\pi^\pm\piz]_D h^\mp) &\propto& 1+r_D^2r_B^2+2r_D r_B\kappa_D\cos(\delta_B-\delta_D\mp\gamma) \nonumber \\
&& -\,\alpha y(1+r_B^2)r_D\kappa_D\cos\delta_D -\alpha y(1+r_D^2)r_B\cos(\delta_B\mp\gamma)\ \ \ \label{rateFav}  \\
&&- \,\alpha x(1-r_B^2)r_D\kappa_D\sin\delta_D+\alpha x(1-r_D^2)r_B\sin(\delta_B\mp\gamma). \nonumber  
\end{eqnarray}


The \CP observables reported in this paper are all experimentally-robust ratios of decay rates. From the ADS modes, two 
ratios of suppressed to favoured decay rates are  measured independently for \Bm and \Bp decays,
\begin{equation}
    \label{eq:rk}
    R_\kaon^\mp \equiv \ratioFrac{\kSUPpm}{\kFAVpm},
\end{equation}
\begin{equation}
    \label{eq:rp}
    R_\pi^\mp \equiv \ratioFrac{\piSUPpm}{\piFAVpm}.
\end{equation}
For the GLW modes two \CP asymmetries are measured,
\begin{eqnarray}
    \Aglwk&=&\frac{\asymmTerm{\BmtoGLWk}{-}{\BptoGLWk}}{\asymmTerm{\BmtoGLWk}{+}{\BptoGLWk}}\ ,\\
    \Aglwpi&=&\frac{\asymmTerm{\BmtoGLWp}{-}{\BptoGLWp}}{\asymmTerm{\BmtoGLWp}{+}{\BptoGLWp}}\
\end{eqnarray}
where $h$ is either a kaon or a pion. Two double ratios are constructed,
\begin{equation}
    \label{eq:Rglw}
    \Rglwkk=\Rkpikk/\Rkpi,\hspace{15mm} \Rglwpipi=\Rkpipipi/\Rkpi,
\end{equation}
where
\begin{equation}
    \label{eq:Rkpihh}
    \Rkpihh=\frac{\asymmTerm{\BmtoGLWk}{+}{\BptoGLWk}}{\asymmTerm{\BmtoGLWp}{+}{\BptoGLWp}}
\end{equation}
is the ratio of the summed-over-charge partial widths for the $\btodk$ decays over the $\btodpi$ decays for a given $D$ meson decay mode.
Last, the \CP asymmetry in the favoured mode is also included, though the expectation from Eq.~\ref{rateFav} is that the \CP asymmetry is only $\mathcal{O}(1\%)$,

\asymmEq{\Akfav}{\kFAVminus}{\kFAVplus}{Akfav}{.}



To summarise, 11 observables are reported: four ADS ratios $R^\mp_{h}$, four GLW asymmetries $\Aglw$, two double ratios $\Rglw$ and the favoured-mode asymmetry, $\Akfav$. 

\section{The \lhcb detector}
\label{sec:detector}
The analysis uses data collected by the \lhcb experiment in proton-proton $(pp)$ collisions at $\sqrt{s}=7\tev$, $8\tev$, and $13\tev$, corresponding to integrated luminosities of $1\invfb$, $2\invfb$, and $6\invfb$ respectively.

The \lhcb detector~\cite{LHCb-DP-2008-001,LHCb-DP-2014-002} is a single-arm forward
spectrometer covering the \mbox{pseudorapidity} range $2<\eta <5$,
designed for the study of particles containing \bquark or \cquark
quarks. The detector includes a high-precision tracking system
consisting of a silicon-strip vertex detector surrounding the $pp$
interaction region, a large-area silicon-strip detector located
upstream of a dipole magnet with a bending power of about
$4{\mathrm{\,Tm}}$, and three stations of silicon-strip detectors and straw
drift tubes placed downstream of the magnet.
The tracking system provides a measurement of the momentum, \ptot, of charged particles with
a relative uncertainty that varies from 0.5\% at low momentum to 1.0\% at 200\gevc.
The minimum distance of a track to a primary $pp$ collision vertex (PV), the impact parameter (IP), 
is measured with a resolution of $(15+29/\pt)\mum$,
where \pt is the component of the momentum transverse to the beam, in\,\gevc.
Different types of charged hadrons are distinguished using information
from two ring-imaging Cherenkov (RICH) detectors. 
Photons, electrons and hadrons are identified by a calorimeter system consisting of
scintillating-pad and preshower detectors, an electromagnetic
and a hadronic calorimeter. 
For the photons used to reconstruct \piz candidates in this analysis, the relative uncertainty on their energy measurement is $\sim6.5\%$. 
Muons are identified by a system composed of alternating layers of iron and multiwire
proportional chambers.
The online event selection is performed by a trigger, which consists of a hardware stage, based on information from the calorimeter and muon
systems, followed by a software stage, which applies a full event reconstruction.

Simulated events of each class of signal decay are used in the analysis.
In the simulation $pp$ collisions are generated using \pythia \cite{Sjostrand:2007gs} with a specific \lhcb configuration. 
Decays of hadrons are described by \evtgen \cite{Lange:2001uf}, in which final-state radiation is generated using \photos \cite{davidson2015photos}.
The interaction of the generated particles with the detector, and its response, are implemented using the \geant toolkit \cite{Allison:2006ve} as described in Ref. \cite{LHCb-PROC-2011-006}.

\section{Event selection} \label{sec:selection}
The study is performed with $\btodh$ candidates, where the neutral $\D$ candidate is reconstructed in a three-body final state composed of two charged tracks and a $\piz$ candidate. These charged tracks and the companion charged track used to reconstruct the \Bm candidate are identified as either a kaon or pion.
The $\piz$ candidate is reconstructed from two photons, as recorded by the electromagnetic calorimeter. 

The mass of the reconstructed $\D$ candidate is required to be within $\pm50\mevcc$ of the known $\Dz$ mass~\cite{PDG2020}.
The mass of the $\piz$ candidate must be within $\pm20\mevcc$ of the known $\piz$ mass~\cite{PDG2020}.
Both of these mass windows correspond to approximately twice the mass resolution of the detector.
The $\btodh$ candidates are required to have a mass in the range \mbox{$5000-5900\mevcc$}.
The \piz candidate must also have \ptot above $1\gevc$ and \pt greater than $0.5\gevc$.
The companion particle is required to satisfy \mbox{$0.5<\pt<10\gevc$} and \mbox{$5<p<100\gevc$}, while the charged decay products from the $\D$ meson must have $\pt>0.25\gevc$.
The mass resolution of the \Bm candidate is improved with a fit \cite{Hulsbergen:2005pu} that constrains the $\D$ candidate to its nominal mass and requires the candidate to point back to the primary $pp$ interaction vertex.
Events are required to have been selected by the trigger in one of two ways: by tracks from the \Bm candidate activating the hadronic calorimeter; or by activity in the rest of the event, independent of the \Bm candidate.

Further background suppression is achieved with a boosted decision tree (BDT) \cite{Breiman,AdaBoost} classifier. 
The BDT classifier is trained using a simulated $\btodh$ signal sample and a sample of combinatorial background taken from data where the \B candidate mass is above $5500\mevcc$. The background sample is subdivided in two parallel procedures to ensure the classifier is not applied to events against which it was trained.
The properties used as input to the BDT training are: the $p$ and $\pt$ of the $D$ candidate, the $\piz$, and the companion particle; the $\chisq$ per degree of freedom for the \Bm candidate vertex fit; the $\chisqip$ of the \Bm and $\D$ candidates, where $\chisqip$ is defined as the difference between the $\chisq$ of the PV reconstructed with and without the particle of interest; the flight distance from the PV for both the \Bm and $\D$ candidates; the angle between a line connecting the particle's decay vertex from the PV and the particle's momentum vector, for both the \Bm and $\D$ candidates; the particle identification (PID) confidence level of both photons constituting the $\piz$ candidate; and the sum of charged-track transverse momenta within a cone surrounding the \Bm candidate direction.

\newcommand{\signi}{s/\sqrt{s+b}}
The selection requirement on the BDT output optimises the metric \mbox{$\signi$}, where $s$ is the expected signal yield in the suppressed \mbox{$\btodk$} ADS mode and $b$ is the combinatorial background level as taken from a fit to the favoured mode. This assumes that the suppressed and favoured modes suffer a comparable level of random $D$ and $h^-$ combinations. 
The expected signal yield is calculated as the yield of the favoured $\btodpi$ mode scaled by the ratio of expected branching fractions while taking into account the relative difference in companion particle PID efficiency.
Since the BDT discriminant includes no variables related to the \D decay products the same BDT requirement is used for the selection of GLW and ADS modes.

\newcommand{\Lnpid}[1]{\ln\mathcal{L}_{#1}}
PID from the RICH detectors is essential to distinguish \mbox{$\btodk$} decays among the more abundant \mbox{$\btodpi$} candidates. The PID algorithm considers the likelihood of the pattern of RICH photons under \kaon and \pion mass hypotheses, $\mathcal{L}_{K,\pi}$.
The companion particle in \mbox{$\btodk$} candidates is required to pass a tight selection requiring a high value of \mbox{$(\Lnpid{\kaon} - \Lnpid{\pi})$}. Candidates failing this selection are treated as \mbox{$\btodpi$} candidates in the mass fit.
Looser but mutually exclusive selection requirements are placed on the kaon and pion from the $\D$ meson decay.

\newcommand{\fdd}{\ensuremath{{\rm FD}_D}\xspace}
Additional restrictions are imposed after the application of the BDT classifier and the PID requirements in order to remove specific sources of background.
Contributions from genuine \Bm meson decays that do not include a $\D$ meson (charmless background) are suppressed by a requirement on the flight distance significance, $\fdd$, defined as the distance between the \D and \Bm meson candidate vertices divided by the uncertainty on this measurement.
A requirement of $\fdd>2$ is applied. 
The relevant branching fractions of \mbox{$\Bm\to hhh\piz$} decays are currently unmeasured and their contribution is estimated by measuring the number of \Bm candidates in the \D-mass sideband regions (defined as \mbox{$1615-1715\mevcc$} and \mbox{$2015-2115\mevcc$}) after the \fdd selection has been applied. 
Charmless background remains in the $\kGLWPi$ sidebands after the \fdd requirement.
The average charmless yield from the lower and upper sidebands is measured and fixed as the charmless background in the signal region fit of this mode.

The suppressed ADS decays, \hSUP, are subject to potential contamination from the GLW modes where one of the charged pions (kaons) from the $\D$ candidate is misidentified as a kaon (pion). 
Simulation demonstrates that such contamination is minimal because a single misidentification among the \D~decay products moves the \D~candidate invariant mass out of the \D-mass window.
The suppressed decays suffer crossfeed in the \D-mass window when a true $\Km\pip$ pair from favoured \hFAV decays is misidentified as a $\pim\Kp$ candidate by the PID system.
This background is reduced by vetoing any suppressed mode candidate whose reconstructed $D$ mass, under the exchange of mass hypotheses between the kaon and charged pion, lies within $\pm30\mevcc$ of the nominal $D$ meson mass. 
The residual contamination is estimated by studying the crossfeed remaining in the \D-mass sidebands, and knowledge of the PID efficiency.
The residual crossfeed, expressed as a fraction of the favoured decay yield, is ($0.85\pm 0.04)\times10^{-4}$, which is about 3\% of the size of the $R_\pi^\mp$ observable.
Finally, after all selection, $4\%$ of events contain more than one \Bm candidate. In these cases a single candidate is selected at random.
\section{Mass fit} \label{sec:fit}
The observables defined in Sec.~\ref{sec:formalism} are determined with a binned maximum-likelihood fit to the mass distribution of the selected \Bm candidates. 
A total of sixteen subsamples are fitted simultaneously: the favoured modes; the suppressed ADS modes; and the two GLW modes, each separated according to the charge of the \Bm candidate, and by the companion track PID. $\btodk$ candidates are defined as those that pass a PID requirement intended to remove $>99\%$ of $\btodpi$ decays; $\btodpi$ candidates are defined as those failing this requirement.
The mass spectra are presented in Figs.~\ref{fig:Bmass_Fig1} to \ref{fig:Bmass_Fig4}.
The total probability density function (PDF) used in the fit is built from five main components, described below, which represent the different categories of signal and background.
\newcommand{\plotBmass}[2]{
\begin{figure}
    \centering
    \includegraphics[height=0.64\textwidth]{figs/#1}
    \caption{#2}
    \label{fig:Bmass_#1}
\end{figure}
}
\plotBmass{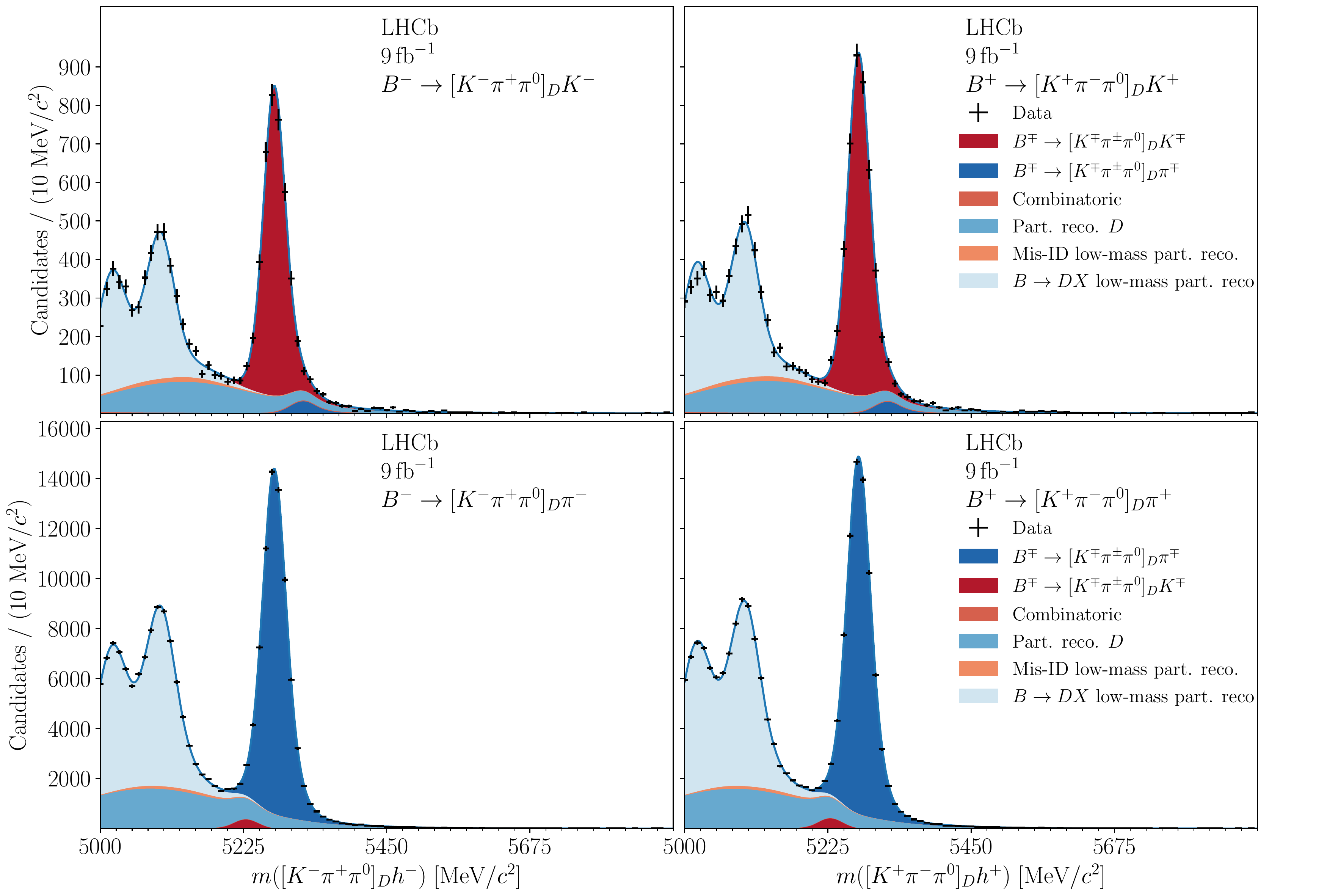}{
Mass distributions of $\Bmp\to[\Kmp\pipm\piz]_D\Kmp$ (top) and $\Bmp\to[\Kmp\pipm\piz]_D\pimp$ (bottom) candidates, separated by charge.
}
\plotBmass{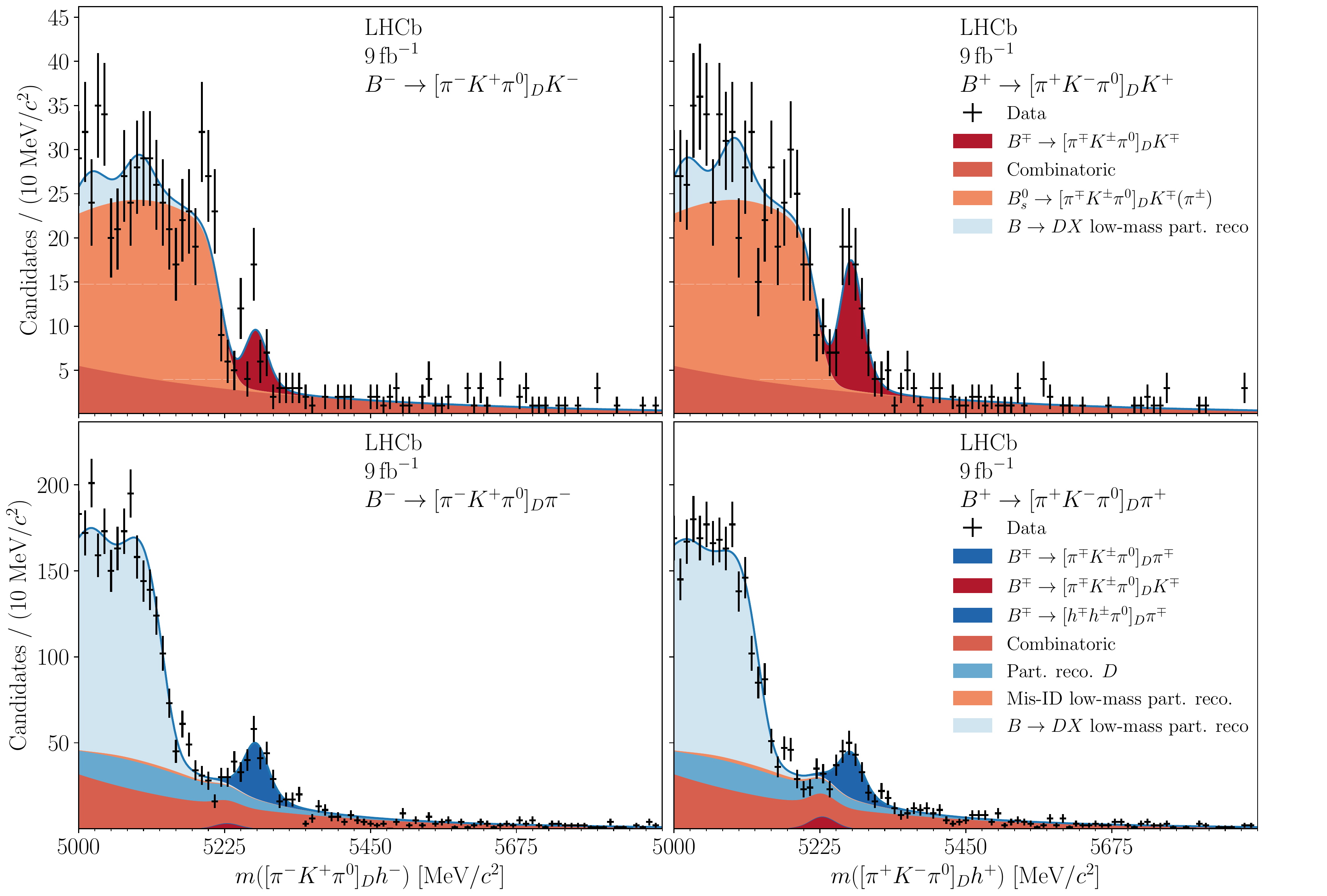}{
Mass distributions of $\Bmp\to[\pimp\Kpm\piz]_D\Kmp$ (top) and $\Bmp\to[\pimp\Kpm\piz]_D\pimp$ (bottom) candidates, separated by charge.
}
\plotBmass{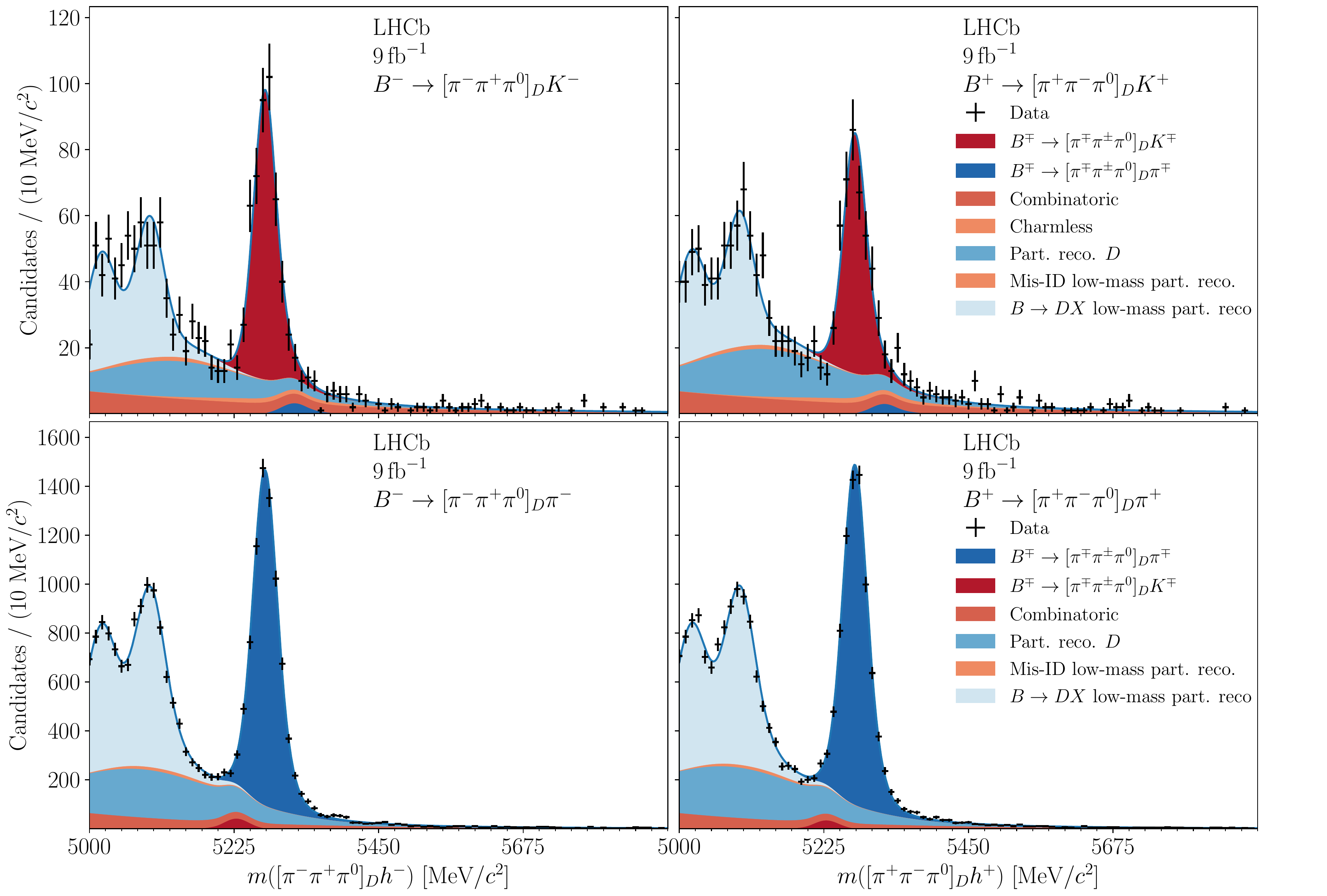}{
Mass distributions of $\Bmp\to[\pipm\pimp\piz]_D\Kmp$ (top) and $\Bmp\to[\pipm\pimp\piz]_D\pimp$ (bottom) candidates, separated by charge.
}
\plotBmass{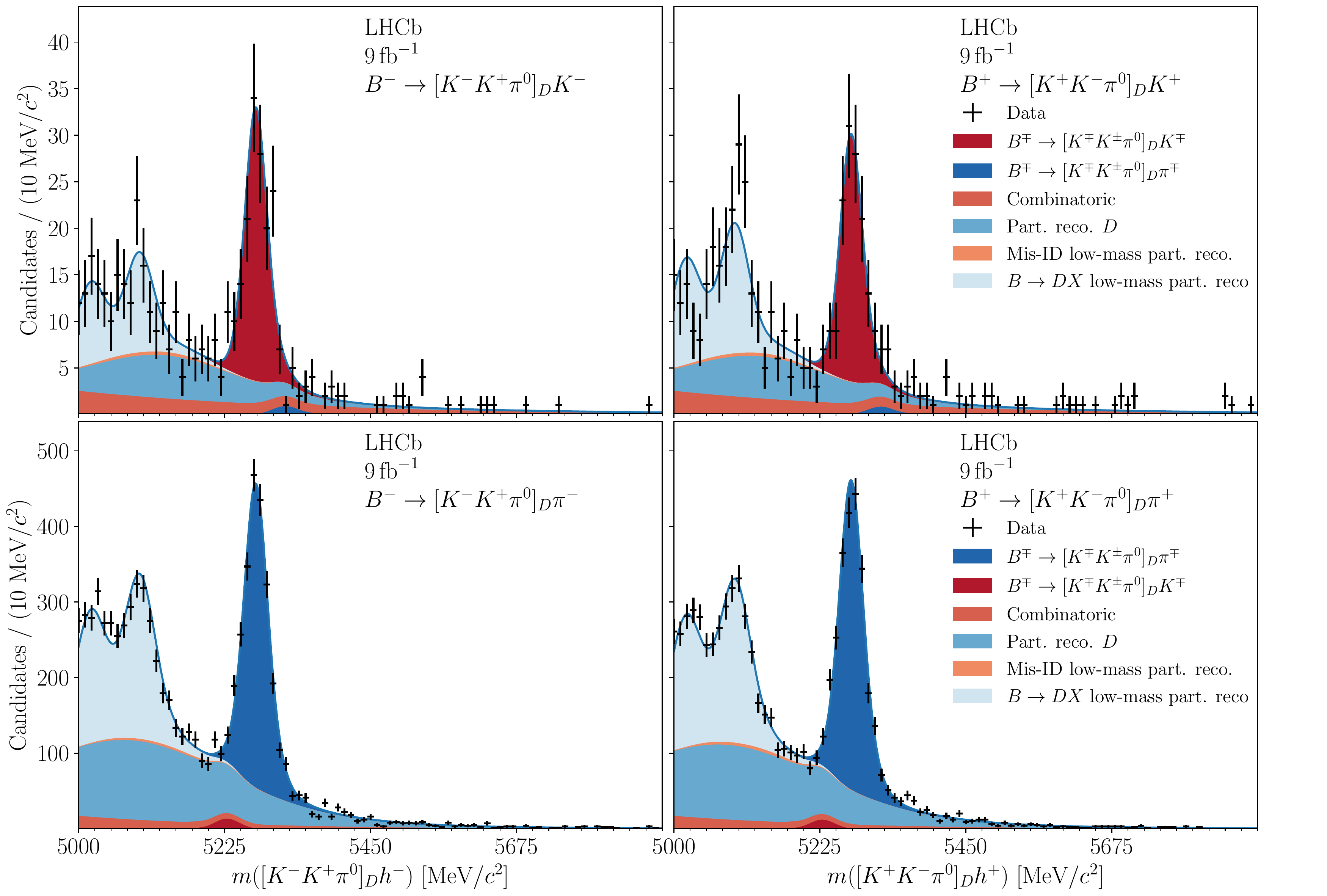}{
Mass distributions of $\Bmp\to[\Kpm\Kmp\piz]_D\Kmp$ (top) and $\Bmp\to[\Kpm\Kmp\piz]_D\pimp$ (bottom) candidates, separated by charge.
}

The mass ($m$) distribution of $\btodpi$ signal candidates is modelled through the use of a modified Gaussian function 
\newcommand{\deltaexcr}{m - \mu}
\begin{equation}
  \label{eq:cruijffextended}
    f(m; \mu, \sigma, \alpha_L, \alpha_R, \beta) = \exp\left(\frac{-(\deltaexcr)^2(1+\beta(\deltaexcr)^2)}{2\sigma^2 + \alpha(\deltaexcr)^2}\right)
\end{equation}
where, $\alpha = \alpha_L$ for $m < \mu$ and $\alpha = \alpha_R$ otherwise.
This expression describes an asymmetric peak of mean $\mu$ and width $\sigma$. The tails of the distribution are parameterised on the left (right) side of the peak by $\alpha_{L(R)}$; the term involving the parameter $\beta$ ensures finite normalisation. All of these parameters vary freely in the fit.
The $\btodk$ signal is modelled using the same modified Gaussian function of Eq. \ref{eq:cruijffextended}. 
All of the parameters are identical to those of the $\btodpi$ modes except for the width, which is related to that of the $\btodpi$ by a ratio parameter that is allowed to vary in the fit.
The $\Bm\to [\kkp]_D\hm$ ($\Bm\to [\ppp]_D\hm$) signal peaks are slightly narrower (wider) than the $\Bm\to [\kpp]_D\hm$ peaks; fixed factors that modify the peak width in these modes are taken from simulation.

$\btodpi$ decays misidentified as $\btodk$ candidates are modelled by the sum of two modified Gaussian functions that share a mean and width but have two sets of tail parameters. 
$\btodk$ decays misidentified as $\btodpi$ candidates are described using a single modified Gaussian function.
Misidentified events form a small background so all their parameters are fixed to the values derived from simulation.

Partially reconstructed $\bquark$-hadron decays populate the mass region below the \Bm mass though their tails can enters the signal region.
Of particular concern are $\Bm(\Bd)$ decays involving a neutral (charged) $\Dstar$ meson decaying to a $\Dz$ candidate plus an unreconstructed neutral (charged) pion.
Similarly, $\Bm\to \Dstarz\hm,\,\Dstarz\to\D\gamma$ decays mimic the signal because the \g is not reconstructed.
There are also contributions from $\Bm(\Bd)$ decays to a $\D$ and a neutral (charged) $\rho$ or $\Kstar$ vector meson decaying to an $\hp\pim(\hp\piz)$ state from which the $\pim(\piz)$ is missed in reconstruction. 
These partially reconstructed decays are described by parabolic functions convolved with a double Gaussian to account for detector resolution~\cite{LHCb-PAPER-2017-021}. 
The yields of these background components vary independently in the fit, with no assumption of \CP symmetry. 
Additionally, partially reconstructed $\Bs\to\Dzb\Km\pip$ decays are an important background for the ADS $\btodk$ signal.
PDFs for this background are determined from simulation weighted with an amplitude model~\cite{LHCb-PAPER-2014-036} and fixed in the fit.
The $\Bs$ yields are allowed to vary freely, but \CP symmetry is assumed because only the $b\to c\overline{u}s$ transition amplitude contributes significantly.

Wrongly reconstructed $\D$ meson decays are a significant source of background under the signal peaks.
These are primarily decays where the $\piz$ candidate is not a decay product of the $\D$ meson, but is wrongly assigned as such.
These contributions are modelled using the modified Gaussian function of Eq.~\ref{eq:cruijffextended} with a large right-hand tail. 
This PDF is defined from an ancillary fit to the \B mass distribution in the $\D$-mass sideband.
The tail and mean parameters are fixed by this procedure but the width is allowed to vary freely in the main fit to account for kinematic differences between the sideband and the signal regions.
The fixed parameters are varied as a source of systematic uncertainty.

The combinatorial background of unassociated \D and \hm candidates is modelled using an exponential function, with a common slope for all \btodk modes and a second for all \btodpi modes.
The favoured and suppressed modes share the same combinatorial background yield.
The GLW modes have independently floating combinatorial yields and \CP symmetry is imposed in all cases.

The observables defined in Section \ref{sec:intro} vary freely in the fit as well as the total \btodpi yields.
The individual signal yields are derived from these values and are presented in Table~\ref{tab:yields}.
The correlation of the statistical uncertainties for the observables are summarised in \mbox{Table \ref{tab:stat_corr}} of Appendix~\ref{app:corrmat}.

\begin{table}[tb]
\centering
\caption{Signal yields for each decay mode. The uncertainties quoted are statistical only.}
\label{tab:yields}
\hspace*{0mm}
\vspace*{0mm}
\begin{tabular}{lr}
\hline
Mode & \multicolumn{1}{c}{Yield}\\ \hline
$B^\pm\rightarrow [K^\pm K^\mp \pi^0]_D\pi^\pm$ & $4026 \pm \phantom{1}77$\\ 
$B^\pm\rightarrow [\pi^\pm \pi^\mp \pi^0]_D\pi^\pm$ & $14180 \pm 140$\\ 
$B^\pm\rightarrow [K^\pm \pi^\mp \pi^0]_D\pi^\pm$ & $140696 \pm 589$\\ 
$B^\pm\rightarrow [\pi^\pm K^\mp \pi^0]_D\pi^\pm$ & $293 \pm \phantom{1}27$\\ 
$B^\pm\rightarrow [K^\pm K^\mp \pi^0]_DK^\pm$ & $401 \pm \phantom{1}29$\\ 
$B^\pm\rightarrow [\pi^\pm \pi^\mp \pi^0]_DK^\pm$ & $1189 \pm \phantom{1}51$\\ 
$B^\pm\rightarrow [K^\pm \pi^\mp \pi^0]_DK^\pm$ & $12265 \pm 158$\\ 
$B^\pm\rightarrow [\pi^\pm K^\mp \pi^0]_DK^\pm$ & $155 \pm \phantom{1}19$\\ 
\hline\end{tabular}
\end{table}



\section{Systematic uncertainties} \label{sec:systematics}
Where fixed parameters are necessary for fit stability, they are varied systematically to assess their contribution to the overall uncertainty. The dominant fixed-parameter effect comes from the wrongly-reconstructed \D-meson PDF, where the systematic variation is defined by the covariance matrix of the ancillary fits to the sideband distributions.

The efficiency and uncertainty of the PID requirements on the companion track are determined from a sample of more than $100$ million $\Dstarpm$ decays reconstructed as $\Dstarpm\to\D\pipm$ with $\D\to\Kmp\pipm$. 
This reconstruction is performed entirely using kinematic variables and provides a high-purity calibration sample of $\Kpm$ and $\pipm$ tracks. 
The PID efficiency varies as a function of track momentum, pseudorapidity, and detector occupancy. 
The average PID efficiency of the signal is determined by reweighting the calibration spectra in these variables to those of the candidates in the favoured mode sample. 
This average PID efficiency is evaluated to be $64.2\%$ and $99.7\%$ for kaons and pions, respectively. 
Systematic uncertainties of $0.7\%$ and $0.1\%$ for companion kaons and companion pions are attributed to the reweighting procedure in the efficiency determination.
These uncertainties are estimated by varying the binning scheme used in the calibration procedure.

Due to their differing interaction lengths, a small negative asymmetry is expected in the detection efficiency of $\Km$ and $\Kp$ mesons. 
The difference between the kaon and pion detection asymmetries is expected to be  $(-0.869\pm 0.165)\%$ and a raw asymmetry of $(-0.17\pm0.10)\%$ is used for pions. 
These numbers are taken from a dedicated study~\cite{Aaij:2017dzw} and also account for any physical asymmetry between the left and right sides of the \lhcb detector. 
There is no systematic uncertainty from the difference in $\Bm$ and $\Bp$ production cross sections because this effect is absorbed into a global asymmetry parameter, dominated by the favoured $\btodpi$ decay, that is free to vary in the fit. All quoted \CP asymmetries are automatically corrected for this value.

The $\Rkpihh$ observables must be corrected for the ratio of efficiencies of $\btodk$ decays relative to $\btodpi$ decays. These ratios quantify the efficiency differences due to the trigger, reconstruction and selection.
They are measured in simulation to be $(92.6\pm2.3)\%$ for the $\hFAV$ and $\hSUP$ modes, $(98.7\pm2.8)\%$ for the $\hGLWPi$ modes, and $(103.7\pm2.9)\%$ for the $\hGLWK$ modes.
The uncertainties listed are based on the finite size of the simulated samples and are large enough to account for any inaccuracies in the simulation.

In order to estimate the systematic uncertainties from the sources described in this section, the fit is performed many times, varying each source by its assigned uncertainty, under the assumption that a Gaussian distribution is appropriate. 
When sources of systematic uncertainty are correlated, \eg the fixed parameters of a PDF, the variations are drawn from a multidimensional Gaussian distribution according to the associated covariance matrix.
The spread of the fit result for each \CP observables is taken as the systematic uncertainty for that quantity. 
These uncertainties are summarised in Table \ref{tab:all_systs}.
The correlations of the systematic uncertainties are recorded in Table \ref{tab:syst_corr} of Appendix~\ref{app:corrmat}.

\begin{table}
\centering
\caption{
Systematic uncertainties on the observables, multiplied by $10^4$. PID refers to the fixed PID efficiencies of the companion tracks. PDFs refers to the uncertainties in fixing parameters in the fit PDF. Sim refers to the use of simulation to calculate relative efficiencies between the $\btodk$ and $\btodpi$ modes. $A_{\rm instr}$ refers to the interaction and detection asymmetries. \D decay refers to the effect of assuming that the distribution of candidates in the \D meson decay phase-space is not sculpted by the selection. The Total column corresponds to the quadrature sum over the five categories.
}
\label{tab:all_systs}
\hspace*{0mm}
\vspace*{0mm}
\begin{tabular}{lrrrrrr}
\hline
{} & PID & PDFs & Sim & $A_{\rm instr}$ & \D decay & Total\\ \hline
$A_{K}^{KK\pi^{0}}$ & 6.9 & 11.7 & 8.2 & 0.1 & 21.4 & 26.6\\ 
$A_{K}^{K\pi\pi^{0}}$ & 7.2 & 1.2 & 2.3 & 16.7 & 1.7 & 18.4\\ 
$A_{K}^{\pi\pi\pi^{0}}$ & 8.2 & 12.2 & 16.2 & 0.1 & 22.2 & 31.1\\ 
$A_{\pi}^{KK\pi^{0}}$ & 1.6 & 1.4 & 1.0 & 16.7 & 0.0 & 16.9\\ 
$A_{\pi}^{\pi\pi\pi^{0}}$ & 1.5 & 0.7 & 1.3 & 16.7 & 0.0 & 16.8\\ 
$R^{KK\pi^{0}}$ & 24.5 & 28.8 & 31.9 & 0.1 & 5.3 & 49.8\\ 
$R^{\pi\pi\pi^{0}}$ & 15.8 & 26.7 & 24.6 & 0.1 & 5.3 & 40.0\\ 
$R^{-}_{K}$ & 0.7 & 1.3 & 0.8 & 0.1 & 3.4 & 3.8\\ 
$R^{-}_{\pi}$ & 0.0 & 0.2 & 0.2 & 0.1 & 0.3 & 0.4\\ 
$R^{+}_{K}$ & 0.8 & 1.1 & 1.3 & 0.3 & 2.3 & 3.0\\ 
$R^{+}_{\pi}$ & 0.0 & 0.2 & 0.2 & 0.1 & 0.4 & 0.5\\ 
\hline\end{tabular}
\end{table}

The values for the coherence factor, average strong-phase differences and \CP-even fraction are reported in Refs~\cite{Malde:2015mha,Nayak:2014tea,LIBBY2014197,Evans:2016tlp,Ablikim:2021cqw,LHCB-PAPER-2015-057} for a uniform acceptance across the three-body phase space of the $\D$ meson decay, which is not the case in this analysis. 
To assess the impact of an imperfect acceptance, studies are performed with amplitude models for each \D mode and an acceptance function derived from simulation.
The study shows that this is the dominant systematic uncertainty for $\Akglwk$, $\Akglwpi$, $R^\pm_K$, and $R^\pm_\pi$ with an effect in the range $(5\sim 26)\%$ of the magnitude of the associated statistical uncertainty.



\newcommand{\rADSk}{\ensuremath{R_{\text{ADS}(\kaon)}}\xspace}
\newcommand{\aADSk}{\ensuremath{A_{\text{ADS}(\kaon)}}\xspace}
\newcommand{\rADSpi}{\ensuremath{R_{\text{ADS}(\pi)}}\xspace}
\newcommand{\aADSpi}{\ensuremath{A_{\text{ADS}(\pi)}}\xspace}
\section{Results} \label{sec:results}
The final results, as determined by the fit and systematic uncertainty assessment, are
\begin{center}
    \centering
    \begin{tabular}{cclclcl}
    $R^{KK\pi^{0}}$ & $=$ & $\phantom{-}1.021$ & $\pm$ & $0.079$ & $\pm$ & $0.005$ \\
    $R^{\pi\pi\pi^{0}}$ & $=$ & $\phantom{-}0.902$ & $\pm$ & $0.041$ & $\pm$ & $0.004$ \\
    $A_{K}^{K\pi\pi^{0}}$ & $=$ & $-0.024$ & $\pm$ & $0.013$ & $\pm$ & $0.002$ \\
    $A_{K}^{KK\pi^{0}}$ & $=$ & $\phantom{-}0.067$ & $\pm$ & $0.073$ & $\pm$ & $0.003$ \\
    $A_{K}^{\pi\pi\pi^{0}}$ & $=$ & $\phantom{-}0.109$ & $\pm$ & $0.043$ & $\pm$ & $0.003$ \\
    $A_{\pi}^{KK\pi^{0}}$ & $=$ & $-0.001$ & $\pm$ & $0.019$ & $\pm$ & $0.002$ \\
    $A_{\pi}^{\pi\pi\pi^{0}}$ & $=$ & $\phantom{-}0.001$ & $\pm$ & $0.010$ & $\pm$ & $0.002$ \\
    $R^{+}_{K}$ & $=$ & $\phantom{-}0.0179$ & $\pm$ & $0.0024$ & $\pm$ & $0.0003$ \\
    $R^{-}_{K}$ & $=$ & $\phantom{-}0.0085$ & $\pm$ & $0.0020$ & $\pm$ & $0.0004$ \\
    $R^{+}_{\pi}$ & $=$ & $\phantom{-}0.00188$ & $\pm$ & $0.00027$ & $\pm$ & $0.00005$ \\
    $R^{-}_{\pi}$ & $=$ & $\phantom{-}0.00227$ & $\pm$ & $0.00028$ & $\pm$ & $0.00004$, \\
    \end{tabular}
\end{center}
where the statistical uncertainties are listed first and the systematic uncertainties second. All results are compatible with, but better than, previous measurements.
The four $R^\pm_\hadron$ observables can be used to calculate 
\begin{center}
    \centering
    \begin{tabular}{cclclcl}
    \rADSk & $=$ & $\phantom{-}0.0127$ & $\pm$ & $0.0016$ & $\pm$ & $0.0002$\\
    \aADSk & $=$ & $-0.38$ & $\pm$ & $0.12$& $\pm$ & $0.02$\\
    \rADSpi & $=$ & $\phantom{-}0.00207$ & $\pm$ & $0.00020$ & $\pm$ & $0.00003$\\
    \aADSpi & $=$ & $\phantom{-}0.069$ & $\pm$ & $0.094$& $\pm$ & $0.016$, \\
    \end{tabular}
\end{center}
\noindent where 
\begin{equation}
R_{{\rm ADS}(h)}=\frac{\Gamma(\Bm\to[\pim\Kp\piz]_\D \hm)+\Gamma(\Bp\to[\pip\Km\piz]_\D \hp)}{\Gamma(\Bm\to[\Km\pip\piz]_\D \hm)+\Gamma(\Bp\to[\Kp\pim\piz]_\D \hp)}
\end{equation}
and
\begin{equation}
A_{{\rm ADS}(h)}=\frac{\Gamma(\Bm\to[\pim\Kp\piz]_\D \hm)-\Gamma(\Bp\to[\pip\Km\piz]_\D \hp)}{\Gamma(\Bm\to[\pim\Kp\piz]_\D \hm)+\Gamma(\Bp\to[\pip\Km\piz]_\D \hp)}.
\end{equation}

A likelihood-ratio test is used to assess the significance of the previously-unobserved \btodk ADS signal~\cite{Wilks:1938dza}.
This is performed by calculating the quantity $\sqrt{-2\ln(\mathcal{L}_b/\mathcal{L}_{s+b})}$ where $\mathcal{L}_b$ and $\mathcal{L}_{s+b}$ are the maximum-likelihood values of the background-only and signal-plus-background hypotheses, respectively. Including systematic uncertainties, a significance of $7.8$ standard deviations ($\sigma$) is found for the decay $\kSUP$.

\section{Interpretation and conclusions} \label{sec:interpretation}
\begin{figure}[]
  \begin{center}
  \includegraphics[width=0.49\linewidth]{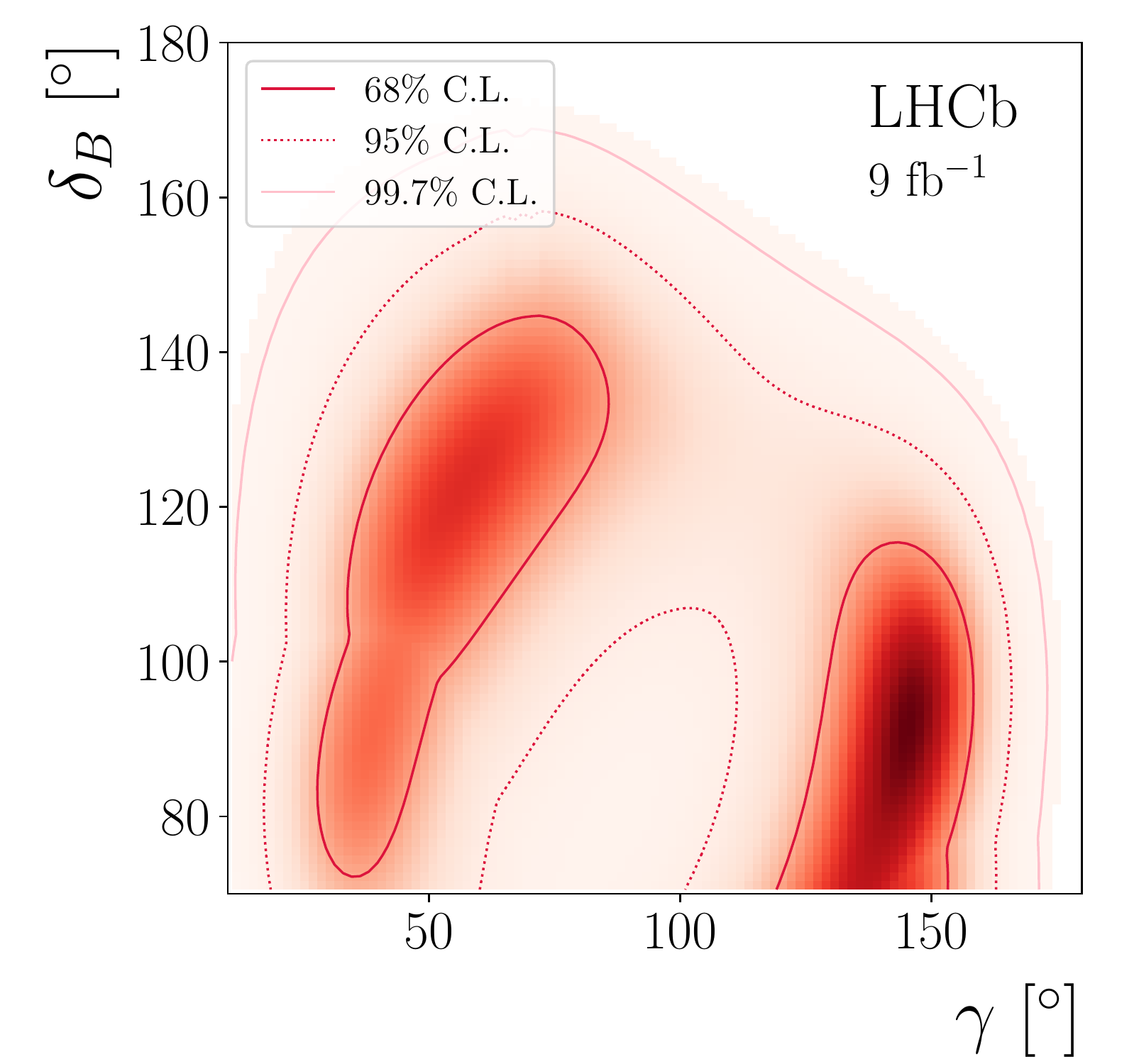}
  \includegraphics[width=0.49\linewidth]{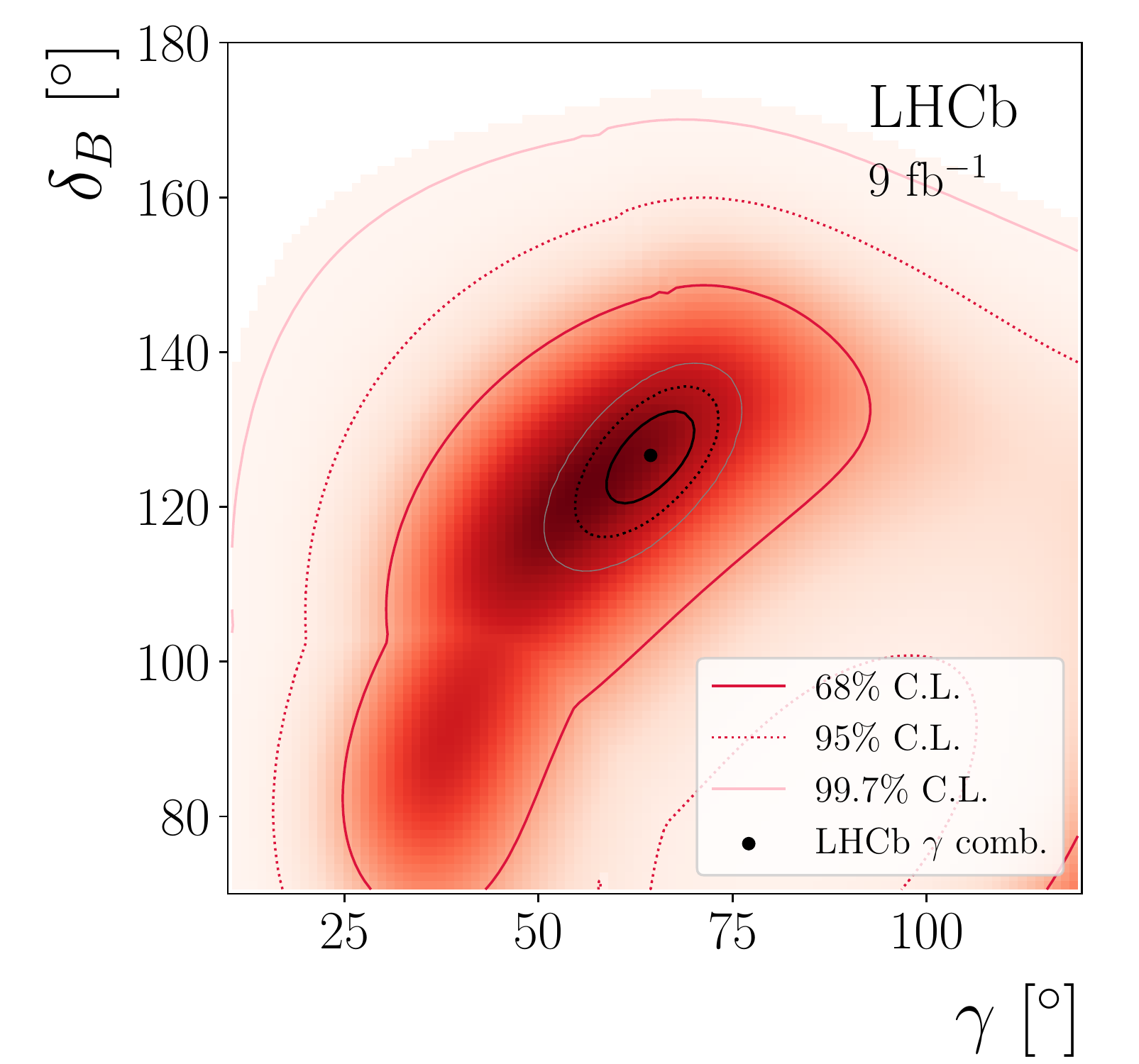}
  \end{center}
  \caption{Confidence regions of the strong phase, $\delta_B$ versus the unitarity triangle angle, \g shows (left) two solutions, one of which is shown (right) to be consistent with the 2021 \g combination result~\cite{LHCb-PAPER-2021-033} whose confidence intervals are superimposed.
  \label{fig:gammadini_ADSGLW_FullReco}}
\end{figure}

The results are interpreted in terms of the fundamental parameters: $\gamma$; $r_B$ and $\delta_B$ using Eqs.~\ref{rateSup}-\ref{rateFav} and inputs from Refs~\cite{Malde:2015mha,Nayak:2014tea,LIBBY2014197,Evans:2016tlp,Ablikim:2021cqw,LHCB-PAPER-2015-057}. 
Confidence intervals are evaluated using the profile likelihood method.
For this, the $\chi^2$ function is evaluated at each point in parameter space to determine a $\Delta \chi^2$ with respect to the best-fit point. Assuming purely Gaussian behaviour, the plotted $p$-value, $p \equiv 1 - \text{CL}$, is given by the probability that $\Delta \chi^2$ is distributed according to a $\chi^2$ distribution with one degree of freedom.
Due to trigonometric ambiguities present in Eq.~\ref{rateSup}, there are up to four solutions in the range $0<\gamma<180^\circ$. The global minimum $\chi^2$ is found at $\gamma=(145^{+\ 9}_{-39})^\circ$ but a second solution, close to the established value\cite{LHCb-PAPER-2021-033}, is also found and quoted below. Two-dimensional confidence regions are shown in Fig.~\ref{fig:gammadini_ADSGLW_FullReco} focusing on the second solution. 
The values of $\gamma$ and the \btodk hadronic parameters, $\delta_B$ and $r_B$, found from this analysis are
\begin{center}
\centering
    \begin{tabular}{lll}
        $\gamma$ & $=$ & $(56^{\,+\,24}_{\,-\,19})^\circ$, \\
        $\delta_B$ & $=$ & $(122^{\,+\,19}_{\,-\,23})^\circ$, \\
        $r_B$ & $=$ & $(9.3^{\,+\,1.0}_{\,-\,0.9})\times10^{-2}$,
    \end{tabular}
\end{center}
with only weak limits found for the \btodpi hadronic parameters.
The corresponding confidence regions for the 2021 LHCb $\gamma$ combination~\cite{LHCb-PAPER-2021-033} are shown for comparison. This new result is consistent with the combination.

In conclusion, using a dataset of $pp$ collisions   corresponding to an integrated luminosity of $9~\invfb$, \btodk and \btodpi decays are studied where the charm meson is reconstructed in the  \mbox{$\Km\pip\piz$}, \mbox{$\pip\pim\piz$}, \mbox{$\Kp\Km\piz$} or \mbox{$\pim\Kp\piz$}  final states.
Eleven \CP observables are measured with world-best precision. The suppressed ${\kSUP}$ mode is observed for the first time, with a significance of $7.8\,\sigma$ and evidence for a large \CP asymmetry in this mode is reported.
The suppressed-to-favoured ratios, $R_{{\rm ADS}(K)}$ and $R_{{\rm ADS}(\pi)}$ are 42 and seven times larger than the doubly-Cabibbo-suppressed \mbox{$\mathcal{B}(\Dz\to\pim\Kp\piz)=(3.05\pm0.15)\times10^{-4}$}~\cite{PDG2020}. In addition to the \CP asymmetry, this underlines the presence and importance of the $b\to u$ transition amplitude contributing to \mbox{$B^\pm\rightarrow [h^\pm h^{\prime\mp}\piz]_Dh^\pm$} decays.  
In combination with similar $B\to DX$ measurements, these results will contribute to a precise determination of the CKM Unitary Triangle angle $\gamma$.

\section*{Acknowledgements}
%
%
\noindent We express our gratitude to our colleagues in the CERN
accelerator departments for the excellent performance of the LHC. We
thank the technical and administrative staff at the LHCb
institutes.
We acknowledge support from CERN and from the national agencies:
CAPES, CNPq, FAPERJ and FINEP (Brazil); 
MOST and NSFC (China); 
CNRS/IN2P3 (France); 
BMBF, DFG and MPG (Germany); 
INFN (Italy); 
NWO (Netherlands); 
MNiSW and NCN (Poland); 
MEN/IFA (Romania); 
MSHE (Russia); 
MICINN (Spain); 
SNSF and SER (Switzerland); 
NASU (Ukraine); 
STFC (United Kingdom); 
DOE NP and NSF (USA).
We acknowledge the computing resources that are provided by CERN, IN2P3
(France), KIT and DESY (Germany), INFN (Italy), SURF (Netherlands),
PIC (Spain), GridPP (United Kingdom), RRCKI and Yandex
LLC (Russia), CSCS (Switzerland), IFIN-HH (Romania), CBPF (Brazil),
PL-GRID (Poland) and NERSC (USA).
We are indebted to the communities behind the multiple open-source
software packages on which we depend.
Individual groups or members have received support from
ARC and ARDC (Australia);
AvH Foundation (Germany);
EPLANET, Marie Sk\l{}odowska-Curie Actions and ERC (European Union);
A*MIDEX, ANR, IPhU and Labex P2IO, and R\'{e}gion Auvergne-Rh\^{o}ne-Alpes (France);
Key Research Program of Frontier Sciences of CAS, CAS PIFI, CAS CCEPP, 
Fundamental Research Funds for the Central Universities, 
and Sci. \& Tech. Program of Guangzhou (China);
RFBR, RSF and Yandex LLC (Russia);
GVA, XuntaGal and GENCAT (Spain);
the Leverhulme Trust, the Royal Society and UKRI (United Kingdom).

\clearpage

\appendix
\section{Correlation matrices} \label{app:corrmat}
The correlation matrix for the statistical uncertainties is presented in Table \ref{tab:stat_corr} and the systematic uncertainty correlation matrix in Table \ref{tab:syst_corr}.

\begin{table}[h!]
\tiny
\centering
\caption{Correlation matrix for statistical uncertainties.}
\label{tab:stat_corr}
\begin{tabular}{c|ccccccccccc}
{} & $R^{KK\pi^{0}}$ & $R^{\pi\pi\pi^{0}}$ & $A_K^{K\pi\pi^{0}}$ & $A_K^{KK\pi^{0}}$ & $A_K^{\pi\pi\pi^{0}}$ & $A_\pi^{KK\pi^{0}}$ & $A_\pi^{\pi\pi\pi^{0}}$ & $R^{+}_{K}$ & $R^{-}_{K}$ & $R^{+}_{\pi}$ & $R^{-}_{\pi}$\\ \hline 
$R^{KK\pi^{0}}$ & $\phantom{-}1.00$ & $\phantom{-}0.05$ & $-0.00$ & $-0.01$ & $-0.00$ & $-0.00$ & $\phantom{-}0.00$ & $\phantom{-}0.02$ & $\phantom{-}0.01$ & $\phantom{-}0.00$ & $\phantom{-}0.00$ \\
$R^{\pi\pi\pi^{0}}$ & {} & $\phantom{-}1.00$ & $-0.00$ & $-0.00$ & $-0.05$ & $-0.00$ & $\phantom{-}0.00$ & $\phantom{-}0.02$ & $\phantom{-}0.01$ & $-0.00$ & $-0.00$ \\
$A_K^{K\pi\pi^{0}}$ & {} & {} & $\phantom{-}1.00$ & $\phantom{-}0.01$ & $\phantom{-}0.02$ & $\phantom{-}0.04$ & $\phantom{-}0.08$ & $-0.01$ & $\phantom{-}0.00$ & $-0.01$ & $\phantom{-}0.01$ \\
$A_K^{KK\pi^{0}}$ & {} & {} & {} & $\phantom{-}1.00$ & $\phantom{-}0.00$ & $-0.01$ & $\phantom{-}0.01$ & $\phantom{-}0.00$ & $\phantom{-}0.00$ & $-0.00$ & $\phantom{-}0.00$ \\
$A_K^{\pi\pi\pi^{0}}$ & {} & {} & {} & {} & $\phantom{-}1.00$ & $\phantom{-}0.01$ & $-0.01$ & $\phantom{-}0.00$ & $\phantom{-}0.00$ & $-0.00$ & $\phantom{-}0.00$ \\
$A_\pi^{KK\pi^{0}}$ & {} & {} & {} & {} & {} & $\phantom{-}1.00$ & $\phantom{-}0.04$ & $-0.00$ & $\phantom{-}0.00$ & $-0.00$ & $\phantom{-}0.00$ \\
$A_\pi^{\pi\pi\pi^{0}}$ & {} & {} & {} & {} & {} & {} & $\phantom{-}1.00$ & $-0.01$ & $\phantom{-}0.00$ & $-0.01$ & $\phantom{-}0.01$ \\
$R^{+}_{K}$ & {} & {} & {} & {} & {} & {} & {} & $\phantom{-}1.00$ & $\phantom{-}0.09$ & $-0.01$ & $\phantom{-}0.00$ \\
$R^{-}_{K}$ & {} & {} & {} & {} & {} & {} & {} & {} & $\phantom{-}1.00$ & $\phantom{-}0.00$ & $-0.01$ \\
$R^{+}_{\pi}$ & {} & {} & {} & {} & {} & {} & {} & {} & {} & $\phantom{-}1.00$ & $\phantom{-}0.01$ \\
$R^{-}_{\pi}$ & {} & {} & {} & {} & {} & {} & {} & {} & {} & {} & $\phantom{-}1.00$\\\hline

\end{tabular}
\end{table}
\begin{table}[h!]
\tiny
\centering
\caption{Correlation matrix for systematic uncertainties.}
\label{tab:syst_corr}
\begin{tabular}{c|ccccccccccc}
{} & $R^{KK\pi^{0}}$ & $R^{\pi\pi\pi^{0}}$ & $A_K^{K\pi\pi^{0}}$ & $A_K^{KK\pi^{0}}$ & $A_K^{\pi\pi\pi^{0}}$ & $A_\pi^{KK\pi^{0}}$ & $A_\pi^{\pi\pi\pi^{0}}$ & $R^{+}_{K}$ & $R^{-}_{K}$ & $R^{+}_{\pi}$ & $R^{-}_{\pi}$\\ \hline 
$R^{KK\pi^{0}}$ & $\phantom{-}1.00$ & $\phantom{-}0.83$ & $-0.29$ & $\phantom{-}0.09$ & $\phantom{-}0.15$ & $-0.23$ & $-0.25$ & $-0.10$ & $-0.05$ & $-0.15$ & $-0.18$ \\
$R^{\pi\pi\pi^{0}}$ & {} & $\phantom{-}1.00$ & $-0.39$ & $-0.07$ & $\phantom{-}0.07$ & $-0.26$ & $-0.28$ & $-0.22$ & $-0.08$ & $-0.15$ & $-0.14$ \\
$A_K^{K\pi\pi^{0}}$ & {} & {} & $\phantom{-}1.00$ & $-0.06$ & $-0.15$ & $\phantom{-}0.91$ & $\phantom{-}0.92$ & $-0.04$ & $-0.12$ & $-0.09$ & $-0.06$ \\
$A_K^{KK\pi^{0}}$ & {} & {} & {} & $\phantom{-}1.00$ & $\phantom{-}0.89$ & $-0.20$ & $-0.19$ & $\phantom{-}0.94$ & $\phantom{-}0.90$ & $\phantom{-}0.66$ & $\phantom{-}0.57$ \\
$A_K^{\pi\pi\pi^{0}}$ & {} & {} & {} & {} & $\phantom{-}1.00$ & $-0.23$ & $-0.23$ & $\phantom{-}0.88$ & $\phantom{-}0.84$ & $\phantom{-}0.54$ & $\phantom{-}0.44$ \\
$A_\pi^{KK\pi^{0}}$ & {} & {} & {} & {} & {} & $\phantom{-}1.00$ & $\phantom{-}1.00$ & $-0.18$ & $-0.17$ & $-0.05$ & $-0.00$ \\
$A_\pi^{\pi\pi\pi^{0}}$ & {} & {} & {} & {} & {} & {} & $\phantom{-}1.00$ & $-0.17$ & $-0.16$ & $-0.06$ & $-0.02$ \\
$R^{+}_{K}$ & {} & {} & {} & {} & {} & {} & {} & $\phantom{-}1.00$ & $\phantom{-}0.92$ & $\phantom{-}0.73$ & $\phantom{-}0.63$ \\
$R^{-}_{K}$ & {} & {} & {} & {} & {} & {} & {} & {} & $\phantom{-}1.00$ & $\phantom{-}0.75$ & $\phantom{-}0.71$ \\
$R^{+}_{\pi}$ & {} & {} & {} & {} & {} & {} & {} & {} & {} & $\phantom{-}1.00$ & $\phantom{-}0.97$ \\
$R^{-}_{\pi}$ & {} & {} & {} & {} & {} & {} & {} & {} & {} & {} & $\phantom{-}1.00$\\\hline

\end{tabular}
\end{table}

\clearpage


\addcontentsline{toc}{section}{References}
\bibliographystyle{LHCb}
\bibliography{main,standard,LHCb-PAPER,LHCb-CONF,LHCb-DP,LHCb-TDR}

\ifx\mcitethebibliography\mciteundefinedmacro
\PackageError{LHCb.bst}{mciteplus.sty has not been loaded}
{This bibstyle requires the use of the mciteplus package.}\fi
\providecommand{\href}[2]{#2}
\begin{mcitethebibliography}{10}
\mciteSetBstSublistMode{n}
\mciteSetBstMaxWidthForm{subitem}{\alph{mcitesubitemcount})}
\mciteSetBstSublistLabelBeginEnd{\mcitemaxwidthsubitemform\space}
{\relax}{\relax}

\bibitem{Brod:2013sga}
J.~Brod and J.~Zupan, \ifthenelse{\boolean{articletitles}}{\emph{{The ultimate
  theoretical error on $\gamma$ from $B \to DK$ decays}},
  }{}\href{https://doi.org/10.1007/JHEP01(2014)051}{JHEP \textbf{01} (2014)
  051}, \href{http://arxiv.org/abs/1308.5663}{{\normalfont\ttfamily
  arXiv:1308.5663}}\relax
\mciteBstWouldAddEndPuncttrue
\mciteSetBstMidEndSepPunct{\mcitedefaultmidpunct}
{\mcitedefaultendpunct}{\mcitedefaultseppunct}\relax
\EndOfBibitem
\bibitem{HFLAV18}
Heavy Flavor Averaging Group, Y.~Amhis {\em et~al.},
  \ifthenelse{\boolean{articletitles}}{\emph{{Averages of $b$-hadron,
  $c$-hadron, and $\tau$-lepton properties as of 2018}},
  }{}\href{https://doi.org/10.1140/epjc/s10052-020-8156-7}{Eur.\ Phys.\ J.\
  \textbf{C81} (2021) 226},
  \href{http://arxiv.org/abs/1909.12524}{{\normalfont\ttfamily
  arXiv:1909.12524}}, {updated results and plots available at
  \href{https://hflav.web.cern.ch}{{\texttt{https://hflav.web.cern.ch}}}}\relax
\mciteBstWouldAddEndPuncttrue
\mciteSetBstMidEndSepPunct{\mcitedefaultmidpunct}
{\mcitedefaultendpunct}{\mcitedefaultseppunct}\relax
\EndOfBibitem
\bibitem{CKMfitter2005}
CKMfitter group, J.~Charles {\em et~al.},
  \ifthenelse{\boolean{articletitles}}{\emph{{\CP violation and the CKM matrix:
  Assessing the impact of the asymmetric $B$ factories}},
  }{}\href{https://doi.org/10.1140/epjc/s2005-02169-1}{Eur.\ Phys.\ J.\
  \textbf{C41} (2005) 1},
  \href{http://arxiv.org/abs/hep-ph/0406184}{{\normalfont\ttfamily
  arXiv:hep-ph/0406184}}, {updated results and plots available at
  \href{http://ckmfitter.in2p3.fr/}{{\texttt{http://ckmfitter.in2p3.fr/}}}}\relax
\mciteBstWouldAddEndPuncttrue
\mciteSetBstMidEndSepPunct{\mcitedefaultmidpunct}
{\mcitedefaultendpunct}{\mcitedefaultseppunct}\relax
\EndOfBibitem
\bibitem{GRONAU1991483}
M.~Gronau and D.~London, \ifthenelse{\boolean{articletitles}}{\emph{{How to
  determine all the angles of the unitarity triangle from $B^0\rightarrow D
  K_S$ and $B_s^0\rightarrow D\phi$}},
  }{}\href{https://doi.org/https://doi.org/10.1016/0370-2693(91)91756-L}{Phys.\
  Lett.\  \textbf{B253} (1991) 483}\relax
\mciteBstWouldAddEndPuncttrue
\mciteSetBstMidEndSepPunct{\mcitedefaultmidpunct}
{\mcitedefaultendpunct}{\mcitedefaultseppunct}\relax
\EndOfBibitem
\bibitem{GRONAU1991172}
M.~Gronau and D.~Wyler, \ifthenelse{\boolean{articletitles}}{\emph{{On
  determining a weak phase from charged $B$ decay asymmetries}},
  }{}\href{https://doi.org/https://doi.org/10.1016/0370-2693(91)90034-N}{Phys.\
  Lett.\  \textbf{B265} (1991) 172}\relax
\mciteBstWouldAddEndPuncttrue
\mciteSetBstMidEndSepPunct{\mcitedefaultmidpunct}
{\mcitedefaultendpunct}{\mcitedefaultseppunct}\relax
\EndOfBibitem
\bibitem{Atwood:1996ci}
D.~Atwood, I.~Dunietz, and A.~Soni,
  \ifthenelse{\boolean{articletitles}}{\emph{{Enhanced CP violation with
  $B\rightarrow KD^0$ modes and extraction of the CKM angle gamma}},
  }{}\href{https://doi.org/10.1103/PhysRevLett.78.3257}{Phys.\ Rev.\ Lett.\
  \textbf{78} (1997) 3257},
  \href{http://arxiv.org/abs/hep-ph/9612433}{{\normalfont\ttfamily
  arXiv:hep-ph/9612433}}\relax
\mciteBstWouldAddEndPuncttrue
\mciteSetBstMidEndSepPunct{\mcitedefaultmidpunct}
{\mcitedefaultendpunct}{\mcitedefaultseppunct}\relax
\EndOfBibitem
\bibitem{PhysRevD.63.036005}
D.~Atwood, I.~Dunietz, and A.~Soni,
  \ifthenelse{\boolean{articletitles}}{\emph{{Improved methods for observing
  $CP$ violation in
  ${B}^{\ifmmode\pm\else\textpm\fi{}}\ensuremath{\rightarrow}{KD}$ and
  measuring the CKM phase $\ensuremath{\gamma}$}},
  }{}\href{https://doi.org/10.1103/PhysRevD.63.036005}{Phys.\ Rev.\ D
  \textbf{63} (2001) 036005}\relax
\mciteBstWouldAddEndPuncttrue
\mciteSetBstMidEndSepPunct{\mcitedefaultmidpunct}
{\mcitedefaultendpunct}{\mcitedefaultseppunct}\relax
\EndOfBibitem
\bibitem{BaBar:2011rud}
BaBar collaboration, J.~P. Lees {\em et~al.},
  \ifthenelse{\boolean{articletitles}}{\emph{{Search for $b \to u$ Transitions
  in $B^\pm \to [K^\mp pi^\pm \pi^0]_D K^\pm$ Decays}},
  }{}\href{https://doi.org/10.1103/PhysRevD.84.012002}{Phys.\ Rev.\ D
  \textbf{84} (2011) 012002},
  \href{http://arxiv.org/abs/1104.4472}{{\normalfont\ttfamily
  arXiv:1104.4472}}\relax
\mciteBstWouldAddEndPuncttrue
\mciteSetBstMidEndSepPunct{\mcitedefaultmidpunct}
{\mcitedefaultendpunct}{\mcitedefaultseppunct}\relax
\EndOfBibitem
\bibitem{Belle:2013dtr}
Belle collaboration, M.~Nayak {\em et~al.},
  \ifthenelse{\boolean{articletitles}}{\emph{{Evidence for the suppressed decay
  $B^-\rightarrow DK^-,\,D\rightarrow K^+\pi^-\pi^0$}},
  }{}\href{https://doi.org/10.1103/PhysRevD.88.091104}{Phys.\ Rev.\ D
  \textbf{88} (2013) 091104},
  \href{http://arxiv.org/abs/1310.1741}{{\normalfont\ttfamily
  arXiv:1310.1741}}\relax
\mciteBstWouldAddEndPuncttrue
\mciteSetBstMidEndSepPunct{\mcitedefaultmidpunct}
{\mcitedefaultendpunct}{\mcitedefaultseppunct}\relax
\EndOfBibitem
\bibitem{LHCb-PAPER-2015-014}
LHCb collaboration, R.~Aaij {\em et~al.},
  \ifthenelse{\boolean{articletitles}}{\emph{{A study of \CP violation in
  \mbox{\decay{\Bmp}{\D h^\mp}} $(h=K,\pi)$ with the modes
  \mbox{\decay{\D}{\Kmp\pipm\piz}}, \mbox{\decay{\D}{\pip\pim\piz}} and
  \mbox{\decay{\D}{\Kp\Km\piz}}}},
  }{}\href{https://doi.org/10.1103/PhysRevD.91.112014}{Phys.\ Rev.\
  \textbf{D91} (2015) 112014},
  \href{http://arxiv.org/abs/1504.05442}{{\normalfont\ttfamily
  arXiv:1504.05442}}\relax
\mciteBstWouldAddEndPuncttrue
\mciteSetBstMidEndSepPunct{\mcitedefaultmidpunct}
{\mcitedefaultendpunct}{\mcitedefaultseppunct}\relax
\EndOfBibitem
\bibitem{Malde:2015mha}
S.~Malde {\em et~al.}, \ifthenelse{\boolean{articletitles}}{\emph{{First
  determination of the $CP$ content of $D \to \pi^+\pi^-\pi^+\pi^-$ and updated
  determination of the $CP$ contents of $D \to \pi^+\pi^-\pi^0$ and $D \to
  K^+K^-\pi^0$}},
  }{}\href{https://doi.org/10.1016/j.physletb.2015.05.043}{Phys.\ Lett.\ B
  \textbf{747} (2015) 9},
  \href{http://arxiv.org/abs/1504.05878}{{\normalfont\ttfamily
  arXiv:1504.05878}}\relax
\mciteBstWouldAddEndPuncttrue
\mciteSetBstMidEndSepPunct{\mcitedefaultmidpunct}
{\mcitedefaultendpunct}{\mcitedefaultseppunct}\relax
\EndOfBibitem
\bibitem{Nayak:2014tea}
M.~Nayak {\em et~al.}, \ifthenelse{\boolean{articletitles}}{\emph{{First
  determination of the CP content of $D \to \pi^+ \pi^- \pi^0$ and $D \to
  K^+K^-\pi^0$}},
  }{}\href{https://doi.org/10.1016/j.physletb.2014.11.022}{Phys.\ Lett.\
  \textbf{B740} (2015) 1},
  \href{http://arxiv.org/abs/1410.3964}{{\normalfont\ttfamily
  arXiv:1410.3964}}\relax
\mciteBstWouldAddEndPuncttrue
\mciteSetBstMidEndSepPunct{\mcitedefaultmidpunct}
{\mcitedefaultendpunct}{\mcitedefaultseppunct}\relax
\EndOfBibitem
\bibitem{Ablikim:2021cqw}
BESIII collaboration, M.~Ablikim {\em et~al.},
  \ifthenelse{\boolean{articletitles}}{\emph{{Measurement of the $D \to
  K^-\pi^+\pi^+\pi^-$ and $D \to K^-\pi^+\pi^0$ coherence factors and average
  strong-phase differences in quantum-correlated ${D\bar{D}}$ decays}},
  }{}\href{https://doi.org/10.1007/JHEP05(2021)164}{JHEP \textbf{05} (2021)
  164}, \href{http://arxiv.org/abs/2103.05988}{{\normalfont\ttfamily
  arXiv:2103.05988}}\relax
\mciteBstWouldAddEndPuncttrue
\mciteSetBstMidEndSepPunct{\mcitedefaultmidpunct}
{\mcitedefaultendpunct}{\mcitedefaultseppunct}\relax
\EndOfBibitem
\bibitem{Rama:2013voa}
M.~Rama, \ifthenelse{\boolean{articletitles}}{\emph{{Effect of $D-\Dbar$ mixing
  in the extraction of $\gamma$ with $B^-\to D^0K^-$ and $B^-\to D\pi^-$
  decays}}, }{}\href{https://doi.org/10.1103/PhysRevD.89.014021}{Phys.\ Rev.\
  \textbf{D89} (2014) 014021},
  \href{http://arxiv.org/abs/1307.4384}{{\normalfont\ttfamily
  arXiv:1307.4384}}\relax
\mciteBstWouldAddEndPuncttrue
\mciteSetBstMidEndSepPunct{\mcitedefaultmidpunct}
{\mcitedefaultendpunct}{\mcitedefaultseppunct}\relax
\EndOfBibitem
\bibitem{LHCb-DP-2008-001}
LHCb collaboration, A.~A. Alves~Jr.\ {\em et~al.},
  \ifthenelse{\boolean{articletitles}}{\emph{{The \lhcb detector at the LHC}},
  }{}\href{https://doi.org/10.1088/1748-0221/3/08/S08005}{JINST \textbf{3}
  (2008) S08005}\relax
\mciteBstWouldAddEndPuncttrue
\mciteSetBstMidEndSepPunct{\mcitedefaultmidpunct}
{\mcitedefaultendpunct}{\mcitedefaultseppunct}\relax
\EndOfBibitem
\bibitem{LHCb-DP-2014-002}
LHCb collaboration, R.~Aaij {\em et~al.},
  \ifthenelse{\boolean{articletitles}}{\emph{{LHCb detector performance}},
  }{}\href{https://doi.org/10.1142/S0217751X15300227}{Int.\ J.\ Mod.\ Phys.\
  \textbf{A30} (2015) 1530022},
  \href{http://arxiv.org/abs/1412.6352}{{\normalfont\ttfamily
  arXiv:1412.6352}}\relax
\mciteBstWouldAddEndPuncttrue
\mciteSetBstMidEndSepPunct{\mcitedefaultmidpunct}
{\mcitedefaultendpunct}{\mcitedefaultseppunct}\relax
\EndOfBibitem
\bibitem{Sjostrand:2007gs}
T.~Sj\"{o}strand, S.~Mrenna, and P.~Skands,
  \ifthenelse{\boolean{articletitles}}{\emph{{A brief introduction to PYTHIA
  8.1}}, }{}\href{https://doi.org/10.1016/j.cpc.2008.01.036}{Comput.\ Phys.\
  Commun.\  \textbf{178} (2008) 852},
  \href{http://arxiv.org/abs/0710.3820}{{\normalfont\ttfamily
  arXiv:0710.3820}}\relax
\mciteBstWouldAddEndPuncttrue
\mciteSetBstMidEndSepPunct{\mcitedefaultmidpunct}
{\mcitedefaultendpunct}{\mcitedefaultseppunct}\relax
\EndOfBibitem
\bibitem{Lange:2001uf}
D.~J. Lange, \ifthenelse{\boolean{articletitles}}{\emph{{The EvtGen particle
  decay simulation package}},
  }{}\href{https://doi.org/10.1016/S0168-9002(01)00089-4}{Nucl.\ Instrum.\
  Meth.\  \textbf{A462} (2001) 152}\relax
\mciteBstWouldAddEndPuncttrue
\mciteSetBstMidEndSepPunct{\mcitedefaultmidpunct}
{\mcitedefaultendpunct}{\mcitedefaultseppunct}\relax
\EndOfBibitem
\bibitem{davidson2015photos}
N.~Davidson, T.~Przedzinski, and Z.~Was,
  \ifthenelse{\boolean{articletitles}}{\emph{{PHOTOS interface in C++:
  Technical and physics documentation}},
  }{}\href{https://doi.org/https://doi.org/10.1016/j.cpc.2015.09.013}{Comp.\
  Phys.\ Comm.\  \textbf{199} (2016) 86},
  \href{http://arxiv.org/abs/1011.0937}{{\normalfont\ttfamily
  arXiv:1011.0937}}\relax
\mciteBstWouldAddEndPuncttrue
\mciteSetBstMidEndSepPunct{\mcitedefaultmidpunct}
{\mcitedefaultendpunct}{\mcitedefaultseppunct}\relax
\EndOfBibitem
\bibitem{Allison:2006ve}
Geant4 collaboration, J.~Allison {\em et~al.},
  \ifthenelse{\boolean{articletitles}}{\emph{{Geant4 developments and
  applications}}, }{}\href{https://doi.org/10.1109/TNS.2006.869826}{IEEE
  Trans.\ Nucl.\ Sci.\  \textbf{53} (2006) 270}\relax
\mciteBstWouldAddEndPuncttrue
\mciteSetBstMidEndSepPunct{\mcitedefaultmidpunct}
{\mcitedefaultendpunct}{\mcitedefaultseppunct}\relax
\EndOfBibitem
\bibitem{LHCb-PROC-2011-006}
M.~Clemencic {\em et~al.}, \ifthenelse{\boolean{articletitles}}{\emph{{The
  \lhcb simulation application, Gauss: Design, evolution and experience}},
  }{}\href{https://doi.org/10.1088/1742-6596/331/3/032023}{J.\ Phys.\ Conf.\
  Ser.\  \textbf{331} (2011) 032023}\relax
\mciteBstWouldAddEndPuncttrue
\mciteSetBstMidEndSepPunct{\mcitedefaultmidpunct}
{\mcitedefaultendpunct}{\mcitedefaultseppunct}\relax
\EndOfBibitem
\bibitem{PDG2020}
Particle Data Group, P.~A. Zyla {\em et~al.},
  \ifthenelse{\boolean{articletitles}}{\emph{{\href{http://pdg.lbl.gov/}{Review
  of particle physics}}}, }{}\href{https://doi.org/10.1093/ptep/ptaa104}{Prog.\
  Theor.\ Exp.\ Phys.\  \textbf{2020} (2020) 083C01}\relax
\mciteBstWouldAddEndPuncttrue
\mciteSetBstMidEndSepPunct{\mcitedefaultmidpunct}
{\mcitedefaultendpunct}{\mcitedefaultseppunct}\relax
\EndOfBibitem
\bibitem{Hulsbergen:2005pu}
W.~D. Hulsbergen, \ifthenelse{\boolean{articletitles}}{\emph{{Decay chain
  fitting with a Kalman filter}},
  }{}\href{https://doi.org/10.1016/j.nima.2005.06.078}{Nucl.\ Instrum.\ Meth.\
  \textbf{A552} (2005) 566},
  \href{http://arxiv.org/abs/physics/0503191}{{\normalfont\ttfamily
  arXiv:physics/0503191}}\relax
\mciteBstWouldAddEndPuncttrue
\mciteSetBstMidEndSepPunct{\mcitedefaultmidpunct}
{\mcitedefaultendpunct}{\mcitedefaultseppunct}\relax
\EndOfBibitem
\bibitem{Breiman}
L.~Breiman, J.~H. Friedman, R.~A. Olshen, and C.~J. Stone, {\em Classification
  and regression trees}, Wadsworth international group, Belmont, California,
  USA, 1984\relax
\mciteBstWouldAddEndPuncttrue
\mciteSetBstMidEndSepPunct{\mcitedefaultmidpunct}
{\mcitedefaultendpunct}{\mcitedefaultseppunct}\relax
\EndOfBibitem
\bibitem{AdaBoost}
Y.~Freund and R.~E. Schapire, \ifthenelse{\boolean{articletitles}}{\emph{A
  decision-theoretic generalization of on-line learning and an application to
  boosting}, }{}\href{https://doi.org/10.1006/jcss.1997.1504}{J.\ Comput.\
  Syst.\ Sci.\  \textbf{55} (1997) 119}\relax
\mciteBstWouldAddEndPuncttrue
\mciteSetBstMidEndSepPunct{\mcitedefaultmidpunct}
{\mcitedefaultendpunct}{\mcitedefaultseppunct}\relax
\EndOfBibitem
\bibitem{LHCb-PAPER-2017-021}
LHCb collaboration, R.~Aaij {\em et~al.},
  \ifthenelse{\boolean{articletitles}}{\emph{{Measurement of \CP observables in
  \mbox{\decay{\Bpm}{D^{(\ast)}\Kpm}} and \mbox{\decay{\Bpm}{D^{(\ast)}\pipm}}
  decays}}, }{}\href{https://doi.org/10.1016/j.physletb.2017.11.070}{Phys.\
  Lett.\  \textbf{B777} (2018) 16},
  \href{http://arxiv.org/abs/1708.06370}{{\normalfont\ttfamily
  arXiv:1708.06370}}\relax
\mciteBstWouldAddEndPuncttrue
\mciteSetBstMidEndSepPunct{\mcitedefaultmidpunct}
{\mcitedefaultendpunct}{\mcitedefaultseppunct}\relax
\EndOfBibitem
\bibitem{LHCb-PAPER-2014-036}
LHCb collaboration, R.~Aaij {\em et~al.},
  \ifthenelse{\boolean{articletitles}}{\emph{{Dalitz plot analysis of
  \mbox{\decay{\Bs}{\Dzb\Km\pip}} decays}},
  }{}\href{https://doi.org/10.1103/PhysRevD.90.072003}{Phys.\ Rev.\
  \textbf{D90} (2014) 072003},
  \href{http://arxiv.org/abs/1407.7712}{{\normalfont\ttfamily
  arXiv:1407.7712}}\relax
\mciteBstWouldAddEndPuncttrue
\mciteSetBstMidEndSepPunct{\mcitedefaultmidpunct}
{\mcitedefaultendpunct}{\mcitedefaultseppunct}\relax
\EndOfBibitem
\bibitem{Aaij:2017dzw}
LHCb collaboration, R.~Aaij {\em et~al.},
  \ifthenelse{\boolean{articletitles}}{\emph{{Measurement of the $B^{\pm}$
  production asymmetry and the $CP$ asymmetry in $B^{\pm} \to J/\psi K^{\pm}$
  decays}}, }{}\href{https://doi.org/10.1103/PhysRevD.95.052005}{Phys.\ Rev.\ D
  \textbf{95} (2017) 052005},
  \href{http://arxiv.org/abs/1701.05501}{{\normalfont\ttfamily
  arXiv:1701.05501}}\relax
\mciteBstWouldAddEndPuncttrue
\mciteSetBstMidEndSepPunct{\mcitedefaultmidpunct}
{\mcitedefaultendpunct}{\mcitedefaultseppunct}\relax
\EndOfBibitem
\bibitem{LIBBY2014197}
J.~Libby {\em et~al.}, \ifthenelse{\boolean{articletitles}}{\emph{{New
  determination of the $D^0 \to K^- \pi^+\pi^0$ and $D^0 \to
  K^-\pi^+\pi^-\pi^+$ coherence factors and average strong-phase differences}},
  }{}\href{https://doi.org/https://doi.org/10.1016/j.physletb.2014.02.032}{Phys.\
  Lett.\  \textbf{B731} (2014) 197}\relax
\mciteBstWouldAddEndPuncttrue
\mciteSetBstMidEndSepPunct{\mcitedefaultmidpunct}
{\mcitedefaultendpunct}{\mcitedefaultseppunct}\relax
\EndOfBibitem
\bibitem{Evans:2016tlp}
T.~Evans {\em et~al.}, \ifthenelse{\boolean{articletitles}}{\emph{{Improved
  determination of the $D \to K^-\pi^+\pi^+\pi^-$ coherence factor and
  associated hadronic parameters from a combination of $e^+e^-\to \psi(3770)\to
  c\bar{c}$ and $pp \to c \bar{c} X$ data}},
  }{}\href{https://doi.org/10.1016/j.physletb.2016.04.037}{Phys.\ Lett.\ B
  \textbf{757} (2016) 520},
  \href{http://arxiv.org/abs/1602.07430}{{\normalfont\ttfamily
  arXiv:1602.07430}}, [Erratum: Phys.Lett.B 765, 402--403 (2017)]\relax
\mciteBstWouldAddEndPuncttrue
\mciteSetBstMidEndSepPunct{\mcitedefaultmidpunct}
{\mcitedefaultendpunct}{\mcitedefaultseppunct}\relax
\EndOfBibitem
\bibitem{LHCB-PAPER-2015-057}
LHCb collaboration, R.~Aaij {\em et~al.},
  \ifthenelse{\boolean{articletitles}}{\emph{{First observation of $\Dz-\Dzb$
  oscillations in \mbox{\decay{\Dz}{\Kp\pip\pim\pim}} decays and a measurement
  of the associated coherence parameters}},
  }{}\href{https://doi.org/10.1103/PhysRevLett.116.241801}{Phys.\ Rev.\ Lett.\
  \textbf{116} (2016) 241801},
  \href{http://arxiv.org/abs/1602.07224}{{\normalfont\ttfamily
  arXiv:1602.07224}}\relax
\mciteBstWouldAddEndPuncttrue
\mciteSetBstMidEndSepPunct{\mcitedefaultmidpunct}
{\mcitedefaultendpunct}{\mcitedefaultseppunct}\relax
\EndOfBibitem
\bibitem{Wilks:1938dza}
S.~S. Wilks, \ifthenelse{\boolean{articletitles}}{\emph{{The large-sample
  distribution of the likelihood ratio for testing composite hypotheses}},
  }{}\href{https://doi.org/10.1214/aoms/1177732360}{Ann.\ Math.\ Stat.\
  \textbf{9} (1938) 60}\relax
\mciteBstWouldAddEndPuncttrue
\mciteSetBstMidEndSepPunct{\mcitedefaultmidpunct}
{\mcitedefaultendpunct}{\mcitedefaultseppunct}\relax
\EndOfBibitem
\bibitem{LHCb-PAPER-2021-033}
LHCb collaboration, R.~Aaij {\em et~al.},
  \ifthenelse{\boolean{articletitles}}{\emph{{Simultaneous determination of
  CKM~angle~$\gamma$ and charm mixing parameters}},
  }{}\href{https://doi.org/10.1007/JHEP12(2021)141}{JHEP \textbf{12} (2021)
  141}, \href{http://arxiv.org/abs/2110.02350}{{\normalfont\ttfamily
  arXiv:2110.02350}}\relax
\mciteBstWouldAddEndPuncttrue
\mciteSetBstMidEndSepPunct{\mcitedefaultmidpunct}
{\mcitedefaultendpunct}{\mcitedefaultseppunct}\relax
\EndOfBibitem
\end{mcitethebibliography}
\newpage
\centerline
{\large\bf LHCb collaboration}
\begin
{flushleft}
\small
R.~Aaij$^{32}$,
A.S.W.~Abdelmotteleb$^{56}$,
C.~Abell{\'a}n~Beteta$^{50}$,
F.J.~Abudinen~Gallego$^{56}$,
T.~Ackernley$^{60}$,
B.~Adeva$^{46}$,
M.~Adinolfi$^{54}$,
H.~Afsharnia$^{9}$,
C.~Agapopoulou$^{13}$,
C.A.~Aidala$^{87}$
% ...

{\footnotesize \it

% ...
$^{9}$Universit{\'e} Clermont Auvergne, CNRS/IN2P3, LPC, Clermont-Ferrand, France\\
% ...
$^{13}$LPNHE, Sorbonne Universit{\'e}, Paris Diderot Sorbonne Paris Cit{\'e}, CNRS/IN2P3, Paris, France\\
% ...
$^{32}$Nikhef National Institute for Subatomic Physics, Amsterdam, Netherlands\\
% ...
$^{46}$Instituto Galego de F{\'\i}sica de Altas Enerx{\'\i}as (IGFAE), Universidade de Santiago de Compostela, Santiago de Compostela, Spain\\
% ...
$^{50}$Physik-Institut, Universit{\"a}t Z{\"u}rich, Z{\"u}rich, Switzerland\\
% ...
$^{54}$H.H. Wills Physics Laboratory, University of Bristol, Bristol, United Kingdom\\
% ...
$^{56}$Department of Physics, University of Warwick, Coventry, United Kingdom\\
% ...
$^{60}$Oliver Lodge Laboratory, University of Liverpool, Liverpool, United Kingdom\\
% ...
$^{68}$Syracuse University, Syracuse, NY, United States\\
% ...
$^{87}$University of Michigan, Ann Arbor, United States, associated to $^{68}$\\
\bigskip
% ...
}
\end{flushleft}
\end{document}